\documentclass[manuscript,11pt,screenauthordraft,nonacm]{acmart}

\AtBeginDocument{%
	\providecommand\BibTeX{{%
			\normalfont B\kern-0.5em{\scshape i\kern-0.25em b}\kern-0.8em\TeX}}}

\usepackage{amsmath,amsfonts}
\usepackage{algorithmic}
\usepackage{graphicx}
\usepackage{hyperref}
\usepackage{listings}
\usepackage{textcomp}
\usepackage[ruled]{algorithm2e}
\usepackage{todonotes}
\usepackage{multirow}
\usepackage{multicol}
\usepackage{caption}
\usepackage{subcaption}

\definecolor{commentGreen}{RGB}{0, 140, 51}
\newcommand \ie {i.e., }
\newcommand \eg {e.g., }
\newcommand{\remove}[1]{}

\newcommand{\varV}{{v}}
\newcommand{\Vars}{{V}}
\newcommand{\dom}{\mathcal{D}}
\newcommand{\values}{\vec{\nu}}

\newcommand{\tuple}[1]{\langle #1 \rangle}
\newcommand{\action}{\sigma}
\newcommand{\actions}{\Sigma}
\newcommand{\prob}{p}
\newcommand{\Sch}{\mathsf{sch} }
\newcommand{ \basicp}{{\mathcal{C}}}
\newcommand{\derp}{{\mathcal{H}}}
\newcommand{\rules}{{\mathcal{R}}}
\newcommand{ \basicpp}{\bar{\basicp}}
\newcommand{\derpp}{\bar{\derp}}
\newcommand{\rulesp}{\bar{{\rules}}}
\newcommand{\pred}{\mathsf{pred}}
\newcommand{\iopred}{\mathsf{doPrequisites}}
\newcommand{\post}{\mathsf{succ}}

\newcommand{\guard}{\phi}
\newcommand{\goal}{\beta}
\newcommand{\Obs}{{\sct{Obs}}}
\newcommand{\game}{\ensuremath{\mathcal{G}}}
\newcommand{\cattacker}{{\semantics{\sattacker}}} 
\newcommand{\sattacker}{\mathrm{AT}} 
\newcommand{\sdefender}{\mathrm{DE}} 
\newcommand{\cdefender}{{\semantics{\sdefender}}} 
\newcommand{\specA}{\mathrm{specA}}

\newcommand{\dist}{\mathcal{P}}
\newcommand{\Dist}{\mathbb{P}}
\newcommand{\opt}{\mathrm{opt}}
\newcommand{\gameTrel}{\Delta}
\newcommand{\smdptrans}{\delta}
\newcommand{\mdptrans}{\overline{\smdptrans}}
\newcommand{\multitransrel}[1]{\overset{#1}{\Longrightarrow} }
\newcommand{\slab}{\mathcal{L}}
\newcommand{\mnext}{\mathsf{X}}
\newcommand{\until}{\mathsf{U}}
\newcommand{\future}{\mathsf{F}}
\newcommand{\strat}{\gamma}
\newcommand{\strats}{\Gamma}
\newcommand{\observable}[1]{[#1]}
\newcommand{\compGame}[2]{#1~||~#2}
\newcommand{\semantics}[1]{{\overline{#1}}}
\newcommand{\MDP}{{\mathrm{M}}}
\newcommand{\defop}{\overset{\mathrm{def}}{=}}

\newcommand{\sat}{\mathsf{Sat}}
\newcommand{\sct}[1]{\ensuremath{ \mathsf{#1}}}
\newcommand{\smallsct}[1]{\ensuremath{{\mathsf{#1}}}}
\newcommand{\coalition}[1]{\langle \! \langle {#1} \rangle \! \rangle}
\newcommand{\reward}{\ensuremath{\xi}\xspace}
\newcommand{\Figname}{Fig.}

\newtheorem{theorem}{Theorem}
\newtheorem{lemma}{Lemma}

\newtheorem{example}{Example}
\newtheorem{definition}{Definition}

 \newcommand{\DefineSnippet}[2]{%
   \expandafter\newcommand\csname snippet--#1\endcsname{%
     \begin{quote}
     \begin{isabelle}
     #2
     \end{isabelle}
     \end{quote}}}
 \newcommand{\Snippet}[1]{\csname snippet--#1\endcsname}
  \DefineSnippet{half_log_simps}{
    @{thm [display] half.simps}
    @{thm_style [display] concl Loga.log.psimps}
 }

\begin{document}
\title{Partially-Observable Security Games for Automating Attack-Defense Analysis
}

\author{Narges Khakpour}
\email{narges.khakpour@ncl.ac.uk}
\affiliation{%
  \institution{School of Computing, Newcastle University}
  \city{Newcastle upon Tyne}
  \country{UK}
}
\affiliation{%
  \institution{Department of Computer Science and Media Technology, Linnaeus University}
  \city{Växjö}
  \country{Sweden}
}
\author{David Parker}
\email{david.parker@cs.ox.ac.uk}
\affiliation{%
  \institution{Department of Computer Science, Oxford University}
  \city{Oxford}
  \country{UK}
}

{%
	\begin{abstract}
Network systems often contain vulnerabilities that remain unfixed in a network for various reasons, such as the lack of a patch or knowledge to fix them. With the presence of such residual vulnerabilities, the network administrator should properly react to the malicious activities or proactively prevent them, by applying suitable countermeasures that minimize the likelihood of an attack by the attacker. In this paper, we propose a stochastic game-theoretic approach for analyzing network security and synthesizing defense strategies to protect a network.  To support analysis under partial observation, where some of the attacker's activities are unobservable or undetectable by the defender, we construct a one-sided partially observable security game and transform it into a perfect game for further analysis. We prove that this transformation is sound for a sub-class of security games and a subset of properties specified in the logic rPATL. We implement a prototype that fully automates our approach and evaluate it by conducting experiments on a real-life network.
\end{abstract}
	
}

\maketitle


%


\section{Introduction}
In today's increasingly complex networked environments, many vulnerabilities still remain in the system after being discovered because of, for example, the lack of a patch or knowledge to fix them, cost factors or organizational preferences for availability and usability over security~\cite{WangILSJ08}. 
Furthermore, hosts and services are frequently added and changed in a complex network, introducing new vulnerabilities and opportunities for  attackers to exploit the system~\cite{Yuan:2014}; a user may connect a new device to or disconnect it from an internal network, bringing new vulnerabilities or facilitating the exploitation of existing ones. In such dynamic and uncertain environments, traditional offline and manual methods to analyze vulnerabilities and defend a network become inadequate. \emph{Enhancing the defense capabilities for network security by employing online and semi-automatic mechanisms has become essential. }

Threat modeling is a common technique used to specify, discover and reason about security weaknesses and vulnerabilities~\cite{ThreatModelingAG,SurveyHONG20171,8017389,MulVALPaper}. It uses models to represent the strategies employed by attackers/defenders to attack/defend a system. 
A wide range of models and languages has been proposed for threat modeling and analysis (see~\cite{SurveyHONG20171,Widel:2019:BFM,ThreatModelingAG} for a survey). 
Attack graph-based ~\cite{sheyner2002automated} and attack tree-based~\cite{weiss1991system} approaches are among the most common ones. Different formal methods have been already used for threat modeling and analysis, e.g. Petri-nets~(see \cite{ThreatModelingAG} for a survey), timed automata~\cite{GadyatskayaHLLO16} and  situation calculus~\cite{samarji2013situation}. \emph{While attack trees have received a lot of attention in the past, the construction of attack trees is often done manually or semi-automatically, which is a serious shortcoming when developing automated online analysis techniques}. {In contrast, attack graphs can be generated fully automatically using formal approaches, such as model checking~\cite{sheyner2002automated} or logic-based reasoning~\cite{MulVALPaper}, which makes them a suitable candidate for automated analysis.}

The opposing objectives and dependent strategies of network attack and defense are very well-suited to game theoretic approaches, which allow us to model the interactions between the attacker and the defender. Game theory has already been applied to analyze network security in different research communities in recent years~\cite{DurkotaLBK15,ChowdharySAHS19,AslanyanNP16} (see \cite{ManshaeiZABH13} for a survey). For instance, Aslayan et al. \cite{AslanyanNP16} propose an approach for formal analysis of attack-defense trees using a stochastic two-player game and probabilistic model checking techniques. \emph{ However, most of the research achievements in this area are based on the hypothesis that the two sides of players have full observation of the opponent's moves, which is unrealistic in practice.} This is because it is not always possible to observe all the attacker's activities: many vulnerabilities and malicious activities cannot be detected due to, for example, the high overhead of monitoring, expensive security controls or ineffective security mechanisms. 

To address the above challenges, we propose a stochastic game-theoretic approach for the analysis of attack-defense scenarios under partial observation to synthesize defense strategies automatically.
We build on stochastic modeling and verification with games~\cite{CFK+13b}, which provide formal assurance about systems that operate in uncertain environments. 
A (turn-based) stochastic multi-player game is a model in which each state is controlled by one of a set of players. In each state, the controlling  player decides on the action to be performed in that state with a probability, i.e. the probability captures the stochastic behavior and uncertainty. A strategy is the choices made by a player to execute an action in each state controlled by that player (based on the history). Strategy synthesis is used to generate a strategy for the players under which the game satisfies a formally specified property.

\begin{figure*}
	\begin{center}
		\includegraphics[scale=0.45]{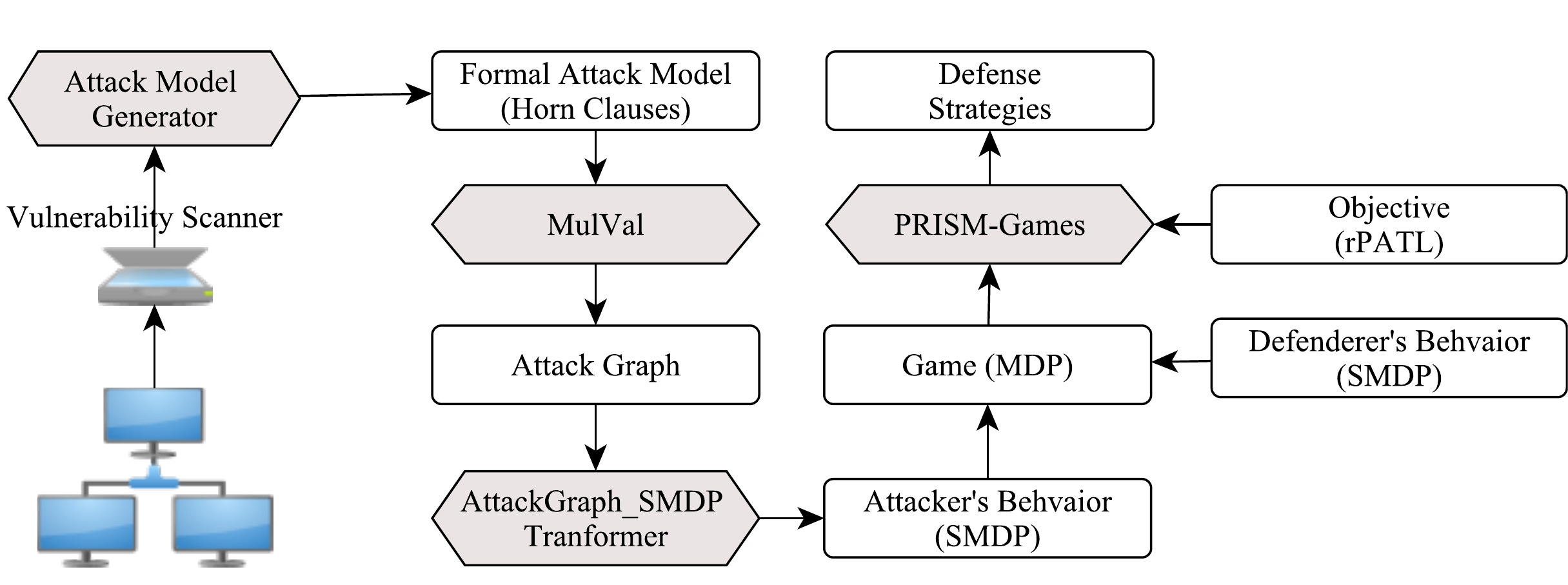}
	\end{center}
	\caption{Our Overall Approach for Automated Attack Analysis and Defense Synthesis}
	\label{fig:prototype}
\end{figure*}

In this paper, we first automatically generate a logical attack graph, which is a type of attack graph that shows different strategies that can be taken by an attacker to reach its goal.
We formalize the vulnerabilities specifications, network scanning results and system configurations automatically using Horn clauses and use MulVAL (Multihost, multistage Vulnerability Analysis)~\cite{MulVALPaper} to generate an attack graph.
We then define the semantics of attack graphs in terms of Symbolic Markov Decision Processes.
A defense is specified using a set of rules, triggered as a reaction to an attack, or a preventive action to stop potential future attacks. A \emph{fully observable} (also called \emph{perfect}) stochastic two-player boolean game is constructed from the semantics of the attacker and the defender's behavior. A perfect game is a game where each of the players has full observation of the other players' actions. We specify the properties to be checked using the temporal logic rPATL (Probabilistic ATL with Rewards)~\cite{CFK+13b} and employ the verification and strategy synthesis techniques for this logic provided by the PRISM-games tool~\cite{KPW18} to analyze a network and synthesize defense strategies. 
Our approach is summarized in Fig.~\ref{fig:prototype}.

To support reasoning under partial observation, we define a partially observable one-sided security game (PO-security game) that provides partial observation of the attacker's behavior. We assume, on the other hand, that the attacker has full observation of the defender's actions. This type of game, where one player has full observation and the other player has partial observation, is called a \emph{one-sided partially observable game}~\cite{Chatterjee0H13}. 
Observe that, if the defender wins in this setting where the game is one-sided, then it will also win in a setting where the attacker has only partial observation \cite{Chatterjee014}.

While the formal verification community has studied analysis under partial observation and presented (theoretical) results, e.g. \cite{ChatterjeeCT13,Chatterjee014,CassezDLLR07,behrmann2007uppaal}, progress on developing practical methods on stochastic games under partial observation has been very limited. We are not aware of any practical method or tool for analyzing partially observable games. 
To be able to analyze a PO-security game, we transform it into a perfect game and use PRISM-games to analyze the perfect security game. We prove that this transformation is sound for a subclass of security games called ODT (Observable Defense Triggers) security games and a subset of rPATL properties (called observable step-unbounded defense objectives), in the sense that a synthesized strategy for the defender in the resulting perfect security game is a valid strategy for the defender in the original PO-security game. 
{We implement a prototype to support and automate the proposed approach and evaluate our approach by applying it to a real-life case study. Our experimental results show that the transformation of a PO-security game improves the performance of our analysis significantly and can also be used as a promising state space reduction technique to handle large attack graphs. }

The contributions of this paper are: (i) automatically constructing a boolean security game model from an attack graph; (ii) introducing partially observable security games and an approach for their efficient analysis; 
(iii) automating our approach and applying it to a real-life case study. 

The paper is organized as follows. Section~\ref{sec:prelimiaries} presents preliminaries on stochastic games. Formalization of attacks graphs is explained in  Section~\ref{sec:attacker.defender.model.specification}. 
Section~\ref{sec:Perfect.Secutiy,game} constructs a fully observable security game 
and Section~\ref{sec::partially.observable.security.game} discusses PO-security games.
Section~\ref{sec:soundness} introduces the soundness theorems of the approach.
We present the evaluation in Section~\ref{sec:evaluation}, discuss related work in Section~\ref{sec:related.work} and conclude in Section~\ref{sec:conlusions}.

\section{Preliminaries}\label{sec:prelimiaries}
In this section, we briefly review probabilistic model checking and strategy synthesis for stochastic games, proposed in \cite{CFK+13b,Sim14}.
We also cover the simpler model of Markov Decision Processes,
needed for our modeling approach.

\subsection{Markov Decision Process}
Let $\Dist_X$ be the set of all discrete probability distribution functions over the set $X$,
$\Vars=\langle \varV_1, \ldots , \varV_n \rangle$ be a vector of variables,
$\dom_{\varV_i}$ be the domain of a variable $\varV_i$,
and $\dom_{\Vars} =  \prod_{i=1}^n \dom_{\varV_i}$. 
A \emph{valuation} $\values$ of $\Vars$ is a tuple
$\langle \values_1, \ldots, \values_n \rangle \in \dom_\Vars$,
and we denote the value of $\varV_i$ in $\values$ by $\values(\varV_i)$, $1 \leq i \leq n$.

\begin{definition}[Symbolic Markov Decision Process] \label{defMDP}
  A \emph{Symbolic Markov Decision Process}  (SMDP) is a tuple $\MDP=
  \tuple{ \Vars,  \values_0, \actions, \smdptrans}$ where 
  $\Vars = \langle  \varV_1, \dots , \varV_n \rangle$ is a tuple of variables,
  $\values_0 \in  \dom_{\Vars}$ gives the initial conditions on the variables, 
	$\actions$ is a finite non-empty set of actions,
  and $\smdptrans$ is a finite set of probabilistic transitions $\tuple{\guard,\action, \dist_U}$ where 
  $\guard\,{\subseteq}\,\dom_{\Vars}$ is a predicate on $\Vars$ which guards the transition, 
   $\action {\in} \actions$ is the transition action,
   {$\dist_U {\in} \Dist_U$ is a discrete probability distribution function over the set $U$} and 
   $U$ is a set of update functions of the form $u : \dom_\Vars \mapsto \dom_{\Vars}$ defined as a set of assignments.
\end{definition}

Initially, $\MDP$ is in its initial state. A transition is fired if its guard is
satisfied and, when fired, the variables are updated with the probabilities corresponding to its update functions.
The semantics of a SMDP is defined in terms of a Markov Decision Process (MDP): 

\begin{definition}[Markov Decision Process] \label{def:MDP}
Let $\MDP= \tuple{ \Vars,  \values_0, \actions, \smdptrans}$ be a SMDP.
The semantics of $\MDP$ is an MDP $\semantics{\MDP}=
  \tuple{V,S, s_0,\actions, \mdptrans}$ where 
  $S\subseteq \dom_\Vars $ is a set of states,
{$s_0 = \values_0$ is the initial state}
  and $\mdptrans: S \times \actions \to \Dist_S$ is a (partial) transition function defined as follows:
  $$\begin{array}{lll}
\mdptrans = \{\langle\langle s, \action \rangle, f(U,s)\rangle~|~\exists \delta=\tuple{\guard,\action, \dist_U} \in \smdptrans, s \models \guard\}\\
f(U,s)=
 \begin{cases} 
      \langle t, \prob \rangle & u \in U, t=u(s), \prob=\dist_U(u) \\
      \textnormal{undef} & \textnormal{otherwise} \\
   \end{cases}
\end{array}
 $$
\end{definition}
\subsection{Stochastic Games}
\begin{definition}[Two-Player Stochastic Game]\label{def:perfect.stochastic.game} 
Let $\semantics{\MDP_1}=\tuple{\Vars_1,S_1, s_{1,0},\actions_1, \mdptrans_1}$ 
and $\semantics{\MDP_2}=\tuple{\Vars_2,S_2, s_{2,0},\actions_2,  \mdptrans_2}$
be two MDPs  specifying two players' behavior.
A (turn-based) stochastic two-player game is a tuple 
$\game= \langle P, V, S, s_0, \actions,  \Sch, \gameTrel \rangle$
where 
$P=\{1,2\}$ is the set of players,
$V$ is a vector of state variables over $V_1 \cup V_2 \cup \{t\}$,
$S \subseteq \dom_{\Vars}$ is the set of states,
$s_0$ is the initial state,
$ \actions = \actions_1 \cup \actions_2$ is the set of actions,
$\Sch:S \to P$ is a scheduler
and $\gameTrel= \{ 
\langle \langle \values, \sigma \rangle, \dist_S \rangle\}$ is a (partial) transition function defined in the below.

Let $\values_{\downarrow_{\Vars_i}}$ be the projection of $\values$ on the variables in $\Vars_i$, $\values \uparrow_{V_i} x$ be the updated $\values$ with $V_i$ to $x$, and $\mdptrans{}(x,\action)(x')$ (similarly for $\gameTrel{}(x,\action)(x')$) represent the probability of a transition from the state $x$ to  the state $x'$ with the action $\action$.
The probability of a transition from $\dist_S$ from the state $\values$ with the active player $i$ to a state $\values'$ by performing the action $\sigma$ is:
$$\begin{array}{lll}
&  
\dist_S(\values') = 
\begin{cases}
p & \values_{\downarrow_{\{t\}}}=i ~~\wedge\phi\\
0& \textnormal{otherwise}
\end{cases}
\end{array}
$$
\textnormal{where} $~\phi = 
\exists x.~\mdptrans_i{}(\values_{\downarrow_{V_i}} ,\action)(x)=p ~\wedge \values' =\values \uparrow_{V_i} x\uparrow_{\{t\}} \Sch(\values).$

\end{definition}
The transition relation in Definition~\ref{def:perfect.stochastic.game} informally states that a state with active player $i$ is updated to a new state $\values'$ in which the active player's state $\values_{\downarrow_{V_i}}$ is updated to $x$ with probability $p$, the other player's local substate remains unmodified and  the active player is updated according to the scheduler function.
We represent the game by $\compGame{\semantics{\MDP_1}}{\semantics{\MDP_2}}$ and partition the game's states into sets $S_1$ and $S_2$, controlled by each of the two players.
Furthermore, $\actions(s)$ denotes the actions that can be performed in state $s \in S$.
We also annotate games with \emph{reward structures} $\reward: \actions\rightarrow\mathbb{R}_{\geq0}$,
which can be used to model rewards or costs.

\subsection{Strategy Synthesis and Model Checking}

Given an objective $\psi$ and a game $\game$, the goal of strategy synthesis is to synthesize  strategies for (coalitions of) players to be able to win the game.
 A strategy informally determines the optimal action that should be performed by the coalition in each state to achieve its objective. 
The objective $\psi$ is specified using the temporal logic rPATL (probabilistic alternating-time temporal logic with rewards).
This logic allows us to express quantitative properties of the games, \eg 
to ensure that the probability of an event occurrence meets some thresholds
(for example: ``can the defender protect the network in a way that the probability of the system being compromised by the attacker is less than 0.1?'').
rPATL is a branching-time logic and contains state formulas $\phi$ and path formulas $\psi$. The syntax is given by the following grammar:
\begin{eqnarray*}
\phi & ::= & \top ~|~  \alpha ~|~ \neg \phi ~|~ \phi \wedge \phi ~|~ \coalition{C} \mathsf{P}_{\bowtie~\!\!p} [\psi] ~|~ \coalition{C} \mathsf{R}^r_{\bowtie~\!\!x} [ \future \phi]
\\
\psi & ::= & \mnext\,\phi ~|~ \phi_1 \until^{\leq k} \phi_2 ~|~ \phi_1 \until \phi_2
\end{eqnarray*}
where $ \alpha$ is a proposition, $C$ is a coalition of players, $\bowtie\,\in \{<, \leq, >, \geq \}$, {$p \in [0,1]$}, $x\in\mathbb{R}$, $k \in \mathbb{N}$ and $r$ is a reward structure.
The operators $\mnext$ ("next"), $\until^{\leq k}$ ("bounded until"), $\until$ ("until") and $\future$ ("eventually") are the standard temporal operators.
Informally, $\coalition{C}\mathsf{P}_{\bowtie\prob} [\psi]$ states that the coalition $C$ has a strategy to ensure that the path formula $\psi$ will be satisfied with a probability meeting the bound $\bowtie\prob$, regardless of the strategies of other players.
Similarly, $\coalition{C}\mathsf{R}_{\bowtie x} [\future \phi]$ means that $C$ has a strategy guaranteeing that the expected reward $r$ accumulated until $\phi$ satisfies $\bowtie x$.

We write $s \models \phi$ to indicate that the state $s$ of $\game$ with the reward structure $r$ satisfies the formula $\phi$, and write $\game \models \phi$, if $s_0 \models \phi$.
Further,  $\sat(\phi) = \{s \in S ~|~ s \models \phi \}$ defines the set of states satisfying $\phi$. 
The semantics of rPATL is defined over games (See Appendix~A
) and it is supported by the probabilistic model checking and strategy synthesis tool {PRISM-games}~\cite{KPW18}.
  The goal of strategy synthesis is to synthesize an optimal strategy for a coalition of players to ensure satisfaction of an objective, regardless of the opponent's strategy.

\section{ Formalized Attack Graphs}\label{sec:attacker.defender.model.specification}

\textbf{Attack Model.}
We use Horn clauses to specify attack models, as done in \cite{MulVALPaper}.
We represent a predicate by $\phi(t_0,\ldots,t_n)$ where a term $t_i$, $0\leq i \leq n$ is either a constant  or a variable.
A rule $r$ defined on predicates is of the form 
$$\phi_0(t_{0,0},\ldots,t_{0,m_0}), \ldots, \phi_k(t_{k,0},\ldots,t_{k,m_k}) \vdash \psi(t'_0,\ldots,t'_n) $$
that informally states, if for all $0\leq j \leq k$, the predicate $\phi_j$ holds, 
then $\psi$ will hold. The left hand side of this rule is called the \emph{body} and $\psi$ is called the \emph{head}.
A rule is safe if and only if the variables occurring in the head also appear in the body. A predicate is \emph{primitive} if it does not appear in the head of any rule with a non-empty body. 
Otherwise, it is called a \emph{derived} predicate. 
A safe rule with an empty body represents a primitive proposition.

 An attack model specifies the system configurations, vulnerabilities and attack actions that lead to exploitation of vulnerabilities. We define an attack model by a triple $spec = \langle  \basicp, \derp , \rules  \rangle $, consisting of
(i) the \emph{primitive capabilities} $\basicp $ that is a set  of primitive predicates, 
(ii) the \emph{derived capabilities} $\derp$ that is a set  of derived predicates and $\derp \cap  \basicp = \emptyset$, 
and (iii) a set of action rules $\rules $ specified in terms of safe rules.

\begin{example}
	The following rule expresses illegal access to a file by exploiting the vulnerability \texttt{{CVE-2018-1000028}}\footnote{https://nvd.nist.gov/vuln/detail/CVE-2018-1000028},
	$$
	\begin{array}{l}
	\sct{accessService(host,fs,ptcl,port)},\\
	\sct{nfsExportInfo(fs, dir, ar, host)},\\
	\sct{execCode(host,service)},\\
	\sct{nfsConfiguration(fs,dir,"rootsquash")},\\
	\sct{vulExists(fs, \texttt{'CVE-2018-1000028'}, mountd)}
	\vdash
	\sct{illegalAccessToFile(fs,write,ex)}.
	\end{array}
	$$
where constants and variables respectively start with a capital and  a small letter.	
This action (a Horn clause) states that if (i) a machine $\sct{host}$ has access to a service provided by a file server $\sct{fs}$ with the protocol $\sct{ptcl}$ on the port $\sct{port}$,
(the primitive predicate $\footnotesize\sct{accessService(host,fs,ptcl,port)} \in \basicp$),
(ii) the file server $\sct{fs}$ exports $\sct{dir}$ with the access right $\sct{ar}$ to the machine $\sct{host}$ with the option $\sct{rootsqush}$ (the predicates \sct{nfsConfiguration} and \sct{nfsExportInfo}),
and (iii) the file server has the vulnerability $\texttt{CVE-2018-1000028}$ (the predicate  \sct{vulExists}),
then the attacker can gain a write access to the file system on the file server (the predicate \sct{illegalAccessToFile}.).
\end{example}

\textbf{Attack Graph.}\label{sec:attack.defense.graph.translation}
An attack goal can vary from compromising a specific machine by the attacker to taking the
whole network down. Given an attack model and a specific attack goal expressed as {a derived predicate}, we generate a graph, called  an attack graph using MulVAL~\cite{MulVALPaper}, that is a reasoning system for modeling and reasoning about the interaction of software bugs with system and network configurations (see Fig.~\ref{fig:prototype}). 

\begin{figure*}
	\centering
	\includegraphics[scale=0.38]{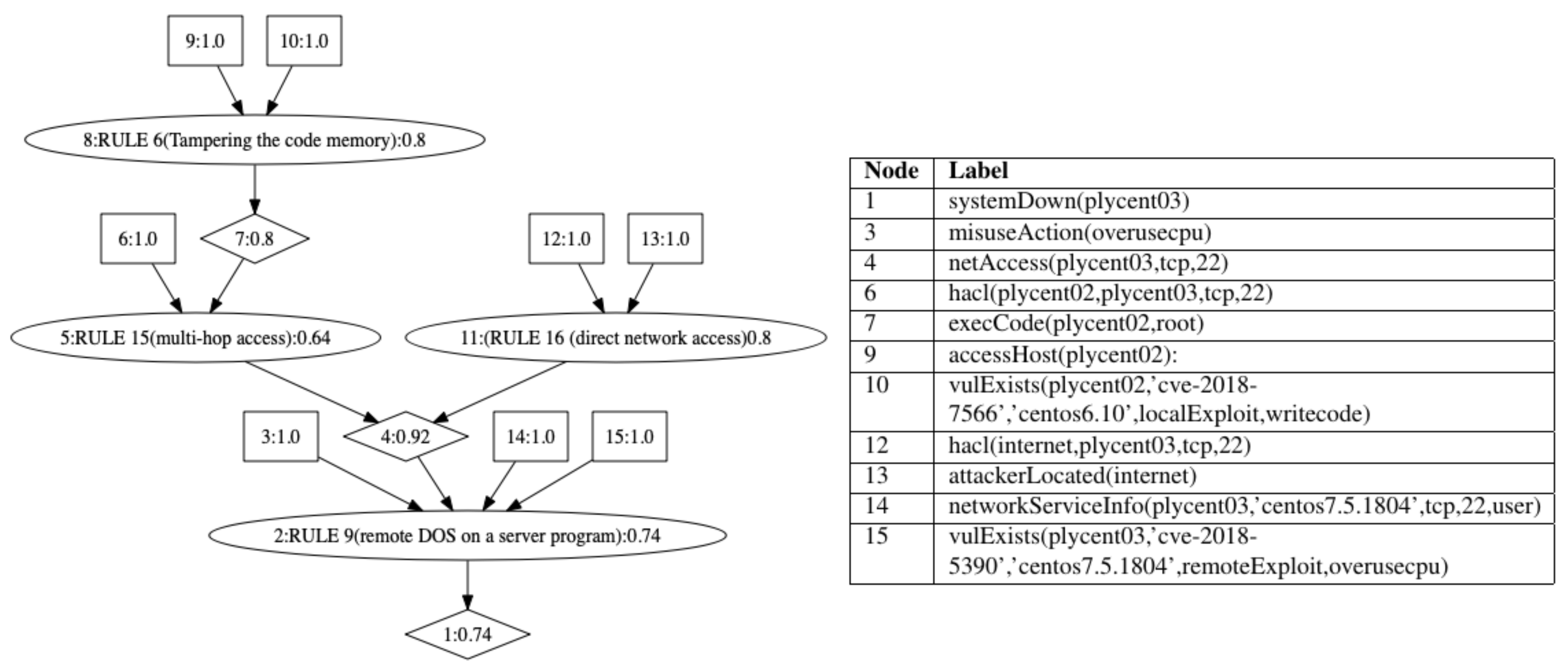}		
	\caption{An Attack Graph}
	\captionsetup{belowskip=-10pt}
	\label{fig:attack.graph}
\end{figure*}

An attack graph shows various strategies that can be taken by an attacker to reach its goal. 
An attack graph consists of three types of nodes:
(a) \emph{condition nodes}, shown by rectangles, that correspond to  primitive capabilities,
(b)  \emph{derived nodes} represented by diamonds and are associated with derived capabilities,
and (c) \emph{ellipse nodes} associated to applying rules in the attack model, \ie a rule node connects a set of condition and/or derived nodes (the rule's body) to a derived node (the rule's head).
Fig.~\ref{fig:attack.graph} shows an attack graph where the table presents the labels of derived and condition nodes as predicates. 
When the prerequisites of a rule in the attack graph are satisfied, it is activated and leads to holding its consequence with a probability.
The label of a node is of the form $(\sct{id},\phi,\prob)$ where $\sct{id}$ is the node identifier and $\prob$ is the probability of reaching that node. If the node is a condition or a derived node, $\phi$ is a proposition, otherwise, it shows the rule name. For readability, we show the labels of the derived and fact nodes in the table.

We rely on the method proposed in ~\cite{WangILSJ08} to calculate the probability of applying an action rule.
This method introduces AGP that is a metric to express the probability that a specific node in the attack graph is reached during an attack. It first associates a score to fact nodes (usually 1.0) and rule nodes that represents the likelihood of an attacker exploiting the exploit, given that all prerequisite conditions are satisfied. We use the scores provided by Common Vulnerability Scoring System (CVSS\footnote{\href{https://www.first.org/cvss/}{https://www.first.org/cvss/}}) to initialize the scores. CVSS is a vulnerability database that provides different scores to measure the severity of vulnerabilities.
This approach, also provides cumulative scores (\ie AGP) that corresponds to the likelihood of an attacker reaching an exploit or condition in the attack graph from an entry point, \ie it considers the causality among attack steps.

\begin{definition}[Attack Graph]
	Let $\goal$ be an attack  goal, $\specA = \langle \basicp, \derp, \rules \rangle $ be an attack model,
	$\basicpp \subseteq \rules$ show the set of {instantiated primitive capabilities} (\ie a set of propositions),
	$\derpp \subseteq \rules$ be the set of {instantiated derived capabilities} in $\rules$,
	and $\bar{\rules}$ be the set of instantiated rules.
	An attack graph is formally defined as a {connected} graph ${G=\langle N, E \rangle}$ where $N=\basicpp \cup \derpp \cup \rulesp$,  $E \subseteq ((\basicpp \cup \derpp)  \times \rulesp) \cup (\rulesp \times \derpp)$,
	and for all $r \in \rulesp$,  
$|\post(r)| = 1$, and 
there is a rule $r' \in \rules$ and a substitution $X$, such that 
		{$\sct{body}(r'[X]) = \underset{\phi_i \in  \pred(r)}{\bigwedge} {\phi_i}$ and $\sct{head}(r'[X])=\post(r)$}
		where $r{[X]}$ instantiates the variables of $r$ according to $X$, the function $\pred(a)$ gives the predecessors of a node (\ie $ \{ \sct{snd}(x)|~ (x,a) \in E \}$ and $\sct{snd(x)}$ returns the second element of the tuple $x$),
		and  $\post(a)$  returns the successors of a node (\ie $ \{ \sct{snd}(x)~|~ (a,x) \in E \}$).
\end{definition}

\section{Fully-Observable Security Games}\label{sec:Perfect.Secutiy,game}
We construct a perfect (fully observable) two-player stochastic security game between the attacker and the defender, where the attacker goal is to gain a specific capability and the defender tries to prevent the attacker from achieving that capability.
In such a turn-based stochastic game, one player independently 
chooses an action in every round and a player's action is fully observable by the opponent,
\ie the attacker and the defender have perfect observation of each other's moves.

\vskip0.6em
\textbf{Security Game Assumptions.}
We make the following assumptions in our security games:
\begin{itemize}
	\item The security game consists of one attacker and a single defender, \ie the attackers attacking the network in a collaborative way are considered as a single attacker.
	\item The attacker's goal is to obtain as many capabilities as possible, and the defender's goal is to prevent the attacker from acquiring more capabilities, and to revoke the attacker's capabilities.  The defender does not attack the attacker. 
	\item If an attacker obtains a capability,  it will hold it until it gets {revoked} by the opponent, \eg if the attacker gains a root privilege on a host, then it will keep that capability until this right is {revoked} by the defender.
	\item An attacker tries an action if it can result in gaining new capabilities, \eg if the attacker obtains access to a service provided by a specific host through exploiting a sequence of vulnerabilities, then they will not try to perform the same sequence of exploitations again, as it will lead to acquiring no new capability.
	\item The defender has the ability to check and observe the consequences of (observable) attacks immediately, \ie we rely on intrusion exploitation systems and intrusion detection systems to detect and report (some of) the attacks immediately.
	
\end{itemize}

\textbf{Attacker Behavior.}
The attacker usually tries various strategies simultaneously,
and if a strategy fails or is defended by the opponent, then she will try a new strategy.
For instance, in the attack graph shown in Fig.~\ref{fig:attack.graph}, the attacker may first try to directly access the http service on the machine $\sct{plycent03}$ to reach this machine (\ie the path via the rule  node 11). However, if the defender blocks the http service (\ie by disallowing the node 12 and blocking the path via the rule node 11), then the attacker may try another strategy by tampering the code memory of the host $\sct{plycent02}$ and access  $\sct{plycent03}$  via $\sct{plycent02}$ (the attack path via the nodes 8 and 5).

To be able to model the changing strategies of the attacker in the semantics, we denote the state of an attacker by the set of its current achieved capabilities.
Given an attack graph, we construct a SDMP using Definition~\ref{def.attack.graph.to.MDP} (\ie \textit{AttackGraph\_SDMP Transformer} module in Fig.~\ref{fig:prototype}).
\begin{definition}[Attacker Behavior]\label{def.attack.graph.to.MDP}
Let $G=\langle \basicpp \cup \derpp \cup \rulesp, E \rangle$ be an attack graph produced from the 
attack model $\specA = \langle \basicp, \derp, \rules \rangle$.
Let ${\prob: \rules \to [0,1]}$ be a {function that denotes the probability of successfully applying a rule $r \in \rules$}.
The semantics of $G$ is defined as the SDMP  $\sattacker= \tuple{\Vars_A, \values_{A,0}, \rulesp, \Delta_A}$ where  {$\Vars_A$ is a vector of boolean variables defined on ${\derpp} \cup  {\basicpp}$,  $\values_{A,0} = \underset{\phi \in \basicpp }{\bigwedge} \phi \wedge\underset{\phi \in {\derpp}}{\bigwedge} \neg \phi $ shows the initial conditions on the variables}, and for each rule $r \in \rulesp$ ,  a transition $\langle \neg e \wedge \underset{\phi_i \in \pred(r)}{\bigwedge} \phi_i, r, \dist_U \rangle  \in \Delta_A$ is defined where $e$ corresponds to the r's successor, and 
  \[
\begin{array}{l}
  U = \{ \emptyset, u\}~,~
  u = \{ e \mapsto \top \}  ~,~\\
  \dist_U(u) = p(r) ~,~ \dist_U(\emptyset) = 1-p(r).
  \end{array}  
\]
\end{definition}

A state variable of an attacker is a boolean variable corresponding to either an instantiated primitive or an instantiated derived capability and indicates whether the attacker holds that capability or not.
Initially the attacker only holds the primitive capabilities in $ \basicpp$ and has no derived capability from $\derpp$.
A transition corresponds to performing an action by the attacker by applying a rule $r$, and when it is triggered (\ie the rule prerequisites hold and its consequence $e$ does not hold), the attacker's gains the new capability $e$ with the probability $p(r)$.
  \begin{example}
  	The attacker's behavior in Fig.~\ref{fig:attack.graph} contains four rule nodes, each is translated to a symbolic transition. For instance, the rule node 2 is translated to the  transition $\langle \phi, \smallsct{remote\_DOS}, \dist_U \rangle$ where 
  	
  		$ \phi = { \begin{array}{l}
  			\smallsct{misuseAction\_overusecpu}\wedge\smallsct{netAccess\_plycent03\_tcp\_22}\wedge\\
	  		\smallsct{networkServiceInfo\_plycent03\_\_centos7\_5} \wedge  \\
  			\smallsct{vulExists\_plycent03\_cve\_2018\_5390\_cen} \wedge\\
  			\neg \smallsct{systemDown\_plycent03}\\
  			\end{array}}$\\\\
$  		 \hspace{0.1in}	
\small  \dist_U(\{\smallsct{systemDown\_plycent03} \mapsto \top\}) = 0.74 \hspace{0.5in} and \hspace{0.5in}
\dist_U(\emptyset) = 0.26.
$\end{example} 
  
  \textbf{Defender Behavior.}\label{sec:attacker.defender.interaction}
   The choices of each player strictly depend on the moves of the other, \ie their capabilities and rights may change as a consequence of the opponent's moves. For instance, if an attacker gains full control over a host as a consequence of a ransomware attack, the defender will have no access to that host anymore. Similarly, if the defender changes the firewall configurations to disallow a service access  on the victim, the attacker may change its strategy to attack the victim.
   Hence, the decision of an attacker to choose a possible attack to an asset should consider the potential countermeasures of the defender to defend it, and vice versa. The countermeasure chosen by the defender to be applied should consider the attacks performed by the attacker and the potential countermeasures that can be applied in future.

  We define two types of defense: \emph{reactive defense} and \emph{proactive defense}. Reactive defenses are usually performed as a reaction to an attack while  a proactive defense is a preventive action performed to prevent future malicious activities.
  {An example of a reactive defense is updating the firewall rules to deny the attacker's access to a service, when a  malicious activity is detected by an intrusion detection system.
  Performing regular vulnerability scans in some critical nodes to detect and fix potential weaknesses, proactively patching applications and monitoring critical security controls are considered as some examples of proactive defenses.} 
 
 The defenses performed by the defender are defined as a reaction to gaining new capabilities by the attacker, to either revoke them or prevent the attacker from achieving specific capabilities in future. 
 
 \begin{definition}[Defender Behavior]\label{def:defener.behavior}
Let $\sattacker= \tuple{\Vars_A, \values_{A,0}, \Sigma_A, \Delta_A}$ be the behavior of an attacker, and $\Vars_{AD} \subseteq \Vars_A$ be the set of defense-triggering capabilities of the attacker. The behavior of the defender is defined as $\sdefender = \tuple{\Vars_D, \values_{D,0}, \actions_D, \Delta_D}$ where $\Vars_{AD} \subseteq \Vars_D$, ${\values_{D,0} }$ implies $ \values_{A,0}$, and $\actions_D \cap \Sigma_A = \emptyset$. 
 \end{definition}
 The defender's state variables includes the defense-triggering state variables of the attacker in addition to some extra internal variables. A defense-triggering variable is a rule consequence that can lead to a reaction by the defender. The actions of the attacker and the defender are disjoint.
  
\begin{example}
A reactive defense to the capability of executing a code with the root privilege by the attacker on the machine $\sct{plycent02}$ is to patch the vulnerability \sct{cve}-\sct{2018}-\sct{7566} (the rule node 8 in Fig.~\ref{fig:attack.graph}). This defense is specified as a transition $\langle \phi, \smallsct{patch\_vul\_2018\_7566}, \dist_U \rangle$ where,

	$ \hspace{-0.1in} { \begin{array}{l}
		\phi = \smallsct{exeCode\_plycent02\_root}\wedge
		\smallsct{vulExists\_plycent02\_vul\_2018\_7566}\\
		u = \{\smallsct{exeCode\_plycent02\_root}\mapsto \bot;\\
		\quad \quad \quad \smallsct{vulExists\_plycent02\_vul\_2018\_7566} \mapsto \bot\}
		\end{array}}$\\
	
	$   \dist_U(u) = 0.85 \hspace{0.5in} and \hspace{0.5in} \dist_U(\emptyset) = 0.15.$
\end{example} 
  	
\textbf{Security Game.}
Given the behavior of the attacker and the defender specified in terms of SDMP and a scheduler $\Sch$, we construct a security game $\game = \langle \{A,D\}, V, S_A \cup S_D, s_0, \actions_D \cup \actions_A,  \Sch, \Delta \rangle $ according to Definition~\ref{def:perfect.stochastic.game} where $\{A,D\}$ is the set of players, $S_A$ is the set of states controlled by the attacker and  $S_D$ is the set of states controlled by the defender (the \emph{Game} module in Fig.~\ref{fig:prototype}). This security game is deterministic, \ie for each action $\action \in \actions$, there is at most one probabilistic transition from each state. 
The security game can be associated with reward structures associating rewards or costs to the states or transitions.

We use the probabilistic model checker PRISM-games to verify various classes of properties for a single player or for both players, e.g., the likelihood or the costs of achieving a specific capability or goal by the attacker. The ultimate goal is to keep the attacker away from its targets where a target can be an intermediate target or a final one, e.g., controlling a specific host and using it as a platform to access the sensitive information on another victim.
PRISM-games can also synthesize a strategy $\strats:S \to \actions$ that identifies an optimal action to be performed in a state by the player controlling that state. We use strategy synthesis to construct defense strategies for protecting the system from attacks.


\section{One-Sided Partially-Observable Security Games}\label{sec::partially.observable.security.game}
In a perfect stochastic game, in every round, one player independently  chooses an action which is fully observable by the opponent, \ie  the player has a full knowledge of the opponent's capabilities. Most security games also make the hypothesis of { perfect observation~\cite{DurkotaLBK15,ChowdharySAHS19}}, \ie the players have full observation of the game state and the opponent's actions, while this is not a realistic assumption in practice.
For instance, the attacker can carry out various attacks in a multi-stage attack to reach its final goal, from a social engineering attack to comprising a host to use as a platform to launch the final attack.
However, not all the attacker's steps are observable by the defender, nor all attacks are trivially detectable, \ie  detecting some of the malicious activities and differentiating them from normal operations can be very hard or expensive.


We assume that the attacker has full observation while the defender has partial observation. This type of games, where one player has full observation and the other player has partial observation, is called \emph{one-sided partially observable game}~\cite{Chatterjee0H13}.  Observe that if the defender wins in this settings where the game is one-sided, then it will win in a setting where the attacker has partial observation \cite{Chatterjee014}. Analogously, if the attacker cannot win in this setting, it cannot win in a setting where the defender has full observation. 
We define a partially observable stochastic two-player security game   as follows:

\begin{definition}[{One-Sided Partially-Observable Security Game}] \label{def:PO.Security.Games}
Let $P= \{A,D\}$, $i\in P$ and $\MDP_i= \tuple{ \Vars_i,  \values_{i,0}, \actions_i, \smdptrans_i}$ be two SMDPs representing the behavior of the attacker and the defender.
A (turn-based) stochastic two-player PO security game is a tuple  $\game= \langle P, \Vars,  S,\Obs, s_0, \actions,  \Sch, \gameTrel{} \rangle$ where  $\Vars$ is a vector defined over $\Vars_A \cup \Vars_D  \cup\{t\}$, $S\subseteq \dom_{\Vars} $ is the set of states and
$\Obs: \dom_{\Vars} \to \dom_{\Vars_D}$ is a function that defines the observation of the defender in a state. The initial  state is {$s_0$ such that $s_0 \models \values_{A,0}$ and $\Obs(s_0) \models \values_{D,0}$},
$\actions = \actions_A \cup \actions_D$ and {$\Sch: S \to P$ is the scheduler}.
The transition $\gameTrel{}: S \times \actions \to \Dist_U$ is a (partial) transition function defined as follows, where {$\sct{up(u,\sigma)}$ is an  update function}:
	$$\begin{array}{lll}
	\gameTrel = \{\langle\langle s, \action \rangle, f(U,s,\sigma)\rangle~|~\exists \delta=\tuple{\guard,\action, \dist_U} \in \smdptrans_A~\wedge~ s \models \guard \vee \\
	\hspace{1in}~\exists \delta=\tuple{\guard,\action, \dist_U} \in \smdptrans_D~\wedge~ \Obs(s) \models \guard\}\\
	f(U,s,\sigma)=
	\begin{cases} 
	\langle  {\sct{up}(u,\sigma)(s)}, \prob \rangle & u \in U, \prob=\dist_U(u) \\
	\textnormal{undef} & \textnormal{otherwise} \\
	\end{cases}
	\end{array}
	$$
\end{definition}

\newcommand{\osattacker}{{O}}

To handle the problem of partial observation in security games, we partition the actions of the attacker into two sets of \emph{observable} and \emph{unobservable} actions and the defender can only observe the observable actions of the attacker and their consequences. An action of the attacker is applying an action rule to obtain a new derived capability; if the action is observable, the defender can also observe the newly obtained capability.
{For instance, the network traffic can be easily monitored to find direct network accesses between hosts (See Fig.~\ref{fig:attack.graph}), while it's impossible to detect a DoS attack if the host is not equipped with a powerful IDS.}
{Therefore, we define the game state as  $V=\langle v_{A,0}, \ldots, v_{A,n},v_{D,0}, \ldots, v_{D,m}, t \rangle$  in this paper where $\Vars_A=\langle v_{A,0}, \ldots, v_{A,n} \rangle$ is the attacker's state, $\Vars_D = \langle v_{D,0}, \ldots, v_{D,m} \rangle$ is the defender's state and $t$ represents the active player. The defender's observation in a game state $\values$ is defined as:
	\begin{equation}\label{eqn:Obs.function}
	\Obs (\values) = \langle \values(v_{D,0}), \ldots, \values(v_{D,m}), t \rangle .
	\end{equation}
	 Note that $\Vars_{AD}$ is duplicated in $\Vars$. Let $\overrightarrow{u}$ be a function that applies the updates of the variables in $\Vars_{AD}$ by $u$ to the sub-states of both players. The function $\sct{up}(u,\sigma)$ applies the update of an unobservable action locally only to the attacker's sub-state. Otherwise, the updates should be applied to the both players' sub-states:  }
 \begin{eqnarray}\label{eqn:up.function}
\sct{up}(u,\sigma)=
\begin{cases} 
u & \sigma \notin\actions_\osattacker \cup\actions_D  \\
\overrightarrow{u}  & \textnormal{otherwise} \\
\end{cases} 
 \end{eqnarray}

\textbf{Analyzing a Partially Observable Security Game.}
To analyze a PO security game, we transform it to a perfect security game and analyze the perfect game instead.
As mentioned above, if a derived capability is obtained due to performing an observable action, it becomes also observable to the defender, otherwise, the update cannot be observed by the defender. If the defender knows the possible consequences of his observation, he can guess the next observable actions by the attacker. To this end, we need to obtain the sets of observable prerequisites of an observable action $\action$, that can lead to performing $\action$ via applying a sequence of unobservable actions, and consequently, achieving an observable derived capability by the attacker.
For instance, assume that the defender cannot detect a memory tampering attack by the attacker, \ie the rule node 8 in Fig.~\ref{fig:attack.graph} becomes unobservable. This means that it will not be possible to detect whether the attacker can run a code with the root privilege on \texttt{plycent02} or not (node 7). However, the defender knows that if the attacker holds the observable capabilities of the nodes 9, 10 and 6, then it can finally perform the action in node 5, \ie the observable prerequisites of the rule 5 will become $\{6,9,10\}$.

There might be several sets of observable prerequisites that can result in performing an observable action. 
For example, let's assume that a multi-hop access by the attacker be unobservable, \ie the action rule node 5 in Fig.~\ref{fig:attack.graph} becomes unobservable. 
The observable prerequisites of the action rule node 2 will be two distinct sets of observable capabilities of $\{14,15,4,3 \}$ and $\{14,15,6,7,3\}$. The node 4 can be achieved either via applying the observable rule 11 (\ie that requires $\{12,13\}$) or via the unobservable rule 5 (\ie that requires $\{6,7\}$).
Since, the node 11 is observable, then the node 4 will become observable too via the path through the node 11; while the same node cannot observed, if the attacker takes the path via the unobservable node 5.

Algorithm~\ref{alg:attack.graph.optimization} returns the \emph{distinct sets of observable prerequisites} (DOP) of a action rule node $r$, where $G=\langle \basicpp \cup \derpp \cup \rulesp, E \rangle$ is an attack graph and, $\actions_\osattacker \subseteq \rulesp$ is the set of observable actions. To obtain the DOP of $r$, it first obtains the DOP of all its prerequisites. If a perquisite is a basic capability, then its DOP contains only itself. If it's a derived capability $n$, its DOP will be the union of the DOPs of all possible rules $r'$ that lead to it immediately (\ie their consequence is $n$). If $r'$ is observable then $\{n\}$ is its only DOP, and, otherwise $r'$'s DOP will be obtained by invoking the algorithm in a recursive way.
{The final $r$'s DOP will be a set of sets where each set contains one element of the DOP from each of r's prerequisite.}
\begin{algorithm}
	\footnotesize
	\caption{$\iopred(r,V)$}
	\label{alg:attack.graph.optimization}
	\KwIn{An action rule  node $r\in \rulesp$ and the visited nodes $V$.}
	\KwOut{Set of distinct observable pre-requisities leading to $r$}
	$\sct{DOPs} = \{\}$\;
	\For{all $n \in  \pred(r)$}{
	\If{$n \in \basicpp$}{$\sct{DOPs}=\sct{DOPs} \cup \{\{n\}\}$\;}
	\Else{ 
		\For{all $r' \in  \pred(n) \backslash V$}{
			\If{$r' \in \actions_\osattacker$}{$ \sct{DOPs} = \sct{DOPs} \cup \{\{n\}\}$\;} 
			\Else{$\sct{DOPs}=\sct{DOPs} \cup \{\iopred(r',V \cup \{r\})\}$\;
					}
	}
	}
}
	\tcc {$\sct{DOPs}=\{D_1, \ldots, D_k\}$: DOPs of $r$'s prerequisites}
	{\Return $\underset{1 \leq i \leq k, e_1 \in D_1, \ldots, e_k \in D_k}{ \bigcup} e_i$	}
\end{algorithm}
\begin{definition}[Attacker's Observable Behavior]\label{def:PO.attack.graph.to.MDP}
	Let $G=\langle \basicpp \cup \derpp \cup \rulesp, E \rangle$ be an attack graph produced from the attack model $\specA = \langle \basicp, \derp, \rules \rangle$.
	Let $\actions_\osattacker \subseteq \rulesp$ be a set of observable actions and ${\prob: \rules \to [0,1]}$ be a {function that returns the probability of successfully applying a rule $r \in \rules$}.
	The observable behavior of $G$ is defined as $\observable{\sattacker}= \tuple{\Vars_\osattacker, \values_{\osattacker,0}, \actions_{\osattacker}, \smdptrans_\osattacker}$ where $\Vars_\osattacker$ is a vector of variables defined on a subset of ${\basicpp} \cup {\derpp}$, 
	$\values_0 = \underset{\phi \in \basicpp}{\bigwedge} \phi \wedge\underset{\phi \in {\derpp \cap \Vars_{\osattacker}}} {\bigwedge} \neg \phi $ shows the initial conditions on the variables. For each rule $r \in \actions_\osattacker$ and each of its distinct set of observable prerequisites $\sct{dop} \in \iopred(r,\{\})$, 
	a transition $\langle \neg e \wedge \underset{\phi \in \sct{dop}}{\bigwedge} \phi, \sct{act(dop)}, \dist_U \rangle  \in \smdptrans_\osattacker$ is defined where $\sct{act(dop)}$ assigns an action label in $\actions_\osattacker$ to $\sct{dop}$, $e$ corresponds to $\sct{succ}(r)$, $U = \{ \emptyset, u\}$, $u = \{ e \mapsto \top \} $, and
	\[
	\begin{array}{l}
	\dist_U(u) = \underset{r_i \in \sct{dop}}{\prod}p(r_i) ,~ \dist_U(\emptyset) = 1- \dist_U(u).
	\end{array}  
	\]
Further, the reward structure for the attacker's actions is defined as $\reward'(r) = \underset{r_i \in \sct{dop}}{\sum} \reward(r_i)$. 
	
\end{definition}

The SMDP in the above definition hides the unobservable actions and only considers {their impact on performing observable actions.} 
For each observable action rule $r$, it finds its DOP, and for each element $\sct{dop}$ of the ODP, adds a new transition whose guard is the conjunction of all the observable capabilities in $\sct{dop}$ as well as the negation of $r$'s consequence.
The action of this transition will set the $r$'s consequence and its probability is the product of all the action rules' probabilities along the path, \ie this path can be taken if all the actions along this path can be performed successfully.

  \begin{example}
  Let all the rule nodes in Fig.~\ref{fig:attack.graph} be unobservable, apart from the rule 2.
  Let $\phi_n$ denote the proposition associated with the node $n$.
  The graph in Fig.~\ref{fig:attack.graph} is translated into two symbolic transitions. 
  One of the transitions is $\langle \phi_{14} \wedge \phi_{15}  \wedge \phi_{12} \wedge \phi_{13}\wedge \phi_3 \wedge \neg \phi_1, \mathtt{path0}, \dist_U \rangle$ where 
  $U= \{\{\phi_1 \mapsto \top\},\emptyset \}$, $\dist_U(\{\phi_1 \mapsto \top\}) = 0.74 * 0.92=0.68$ and $\dist_U(\emptyset) = 0.32$. The other one is defined similarly.
    \end{example}

We use the attacker's behavior constructed in Definition~\ref{def:PO.attack.graph.to.MDP} to build a fully-observable security game and analyze it instead of directly analyzing the PO security games constructed using the functions $\Obs$ and $\sct{up}$ defined in Equation \ref{eqn:Obs.function} and Equation \ref{eqn:up.function}. 
{Let $\Vars'_{AD}$ be the attacker's copy of $\Vars_{AD}$ (\ie the set of common variables of the attacker and the defender). The scheduler of the perfect game is defined as follows where $\Sch$ is the scheduler of the original PO security game:}
\begin{equation}\label{eqn:PO.scheduler.function}
\Sch'(\values) = \Sch(\values_{\downarrow_{(\Vars_{\osattacker} \backslash \Vars'_{AD}) \cup \Vars_D \cup \{t\}}}).
\end{equation}
The scheduler of a state $\values$ is defined as the value returned by the original scheduler $\Sch$ for the corresponding state of~$\values$ in the original game, that is basically a projection on the state variables {of the attacker,} the defender and the variable $t$.


\section{Soundness}\label{sec:soundness}
We define an Observable Defense Triggering  security game (\emph{ODT security game}) as the following:
\begin{definition}\label{def:ODT.security.games}
An \emph{ODT security game} 
is a PO security game in which all the defense-triggering capabilities of the attacker are observable, \ie $V_{AD} \subseteq V_\osattacker$.
\end{definition}
We prove that the transformation of an ODT security game, constructed using Definition~\ref{def:PO.Security.Games} and the functions $\Obs$ and $\sct{up}$ defined in Equation \ref{eqn:Obs.function} and Equation \ref{eqn:up.function} to a perfect security is sound. {
	Since some of the attackers actions are aggregated into one action, some information about the  attacker behavior will be missing in the ODT game. For instance, the number of steps taken by an attacker in the original game cannot be the same as that of ODT game. Hence, it's not always possible to reason about properties that include the number of steps. The same goes for the operator $X$, as a next state in the original game cannot  always be mapped to a state in the ODT game.}
\begin{definition}\label{def:assumptions}
	A  formula $\phi$ is an {observable step-unbounded defense} objective, if (i) it is defined over $\Vars_\osattacker \cup \Vars_{D}$ (\ie it contains no unobservable variable), (ii) it contains no $\mnext${ or $\until^k$  operators, }
	and (iii) in any sub-formula ${\ll{C}\gg} \mathsf{P}_{\bowtie~p} [\psi] $, $C \subseteq \{D\}$. 
\end{definition}

We prove that an observable step-unbounded defense objective $\phi$ holds by an ODT security game defined in Definition~\ref{def:PO.Security.Games}, if and only if it is satisfied by a perfect game with the attacker behavior obtained using Definition~\ref{def:PO.attack.graph.to.MDP}.

\begin{theorem}\label{thm:game.property.verification}
	Let $\game$ be a one-sided partially observable security game and  $\observable{\game}$ be its transformation to a perfect game.
	{For any observable step-unbounded defense objective $\phi$, {$\game \models \phi $ if and only if $\observable{\game} \models \phi $.}
	}
\end{theorem}
Proof.	See Appendix~B.
%
We prove that PRISM-games will synthesize the same strategies for the defender in $\game$ and $\observable{\game}$:
\begin{theorem}\label{thm:game.property.synthesis}
	{For an observable step-unbounded defense objective $\ll D \gg \mathsf{P}_{\bowtie q}[\psi]$ or $\ll D \gg \mathsf{R}_{\bowtie q}[\future\phi]$, PRISM-games synthesizes the same winning strategies for the defender in $\game$ and $\observable{\game}$.
	}
\end{theorem}
Proof.	See Appendix~C.

\newcommand{\ms}[1]{\ensuremath{\mathsf {#1}} \xspace}
\newcommand{\aCosts}{\ensuremath{\mathsf {aCosts}}}
\newcommand{\dCosts}{\ensuremath{\mathsf {dCosts}}}
\newcommand{\attackerBlocked}{\ensuremath{\mathsf {attackerBlocked}} \xspace}
\section{Evaluation}\label{sec:evaluation}
\subsection{Implementation}
We implemented a prototype  in Java and Python to support our approach whose architecture is shown in \Figname~\ref{fig:prototype}. This prototype automatically (i) generates an attack model from the network security scanning results, (ii) generates its attack graph using MulVAL, and (iii) transforms it to an SMDP (the PRISM-games input format). The defender's behavior is generated in a semi-automatic way. Given an rPATL formula and the formalized behavior of the attacker and defender, the tool uses PRISM-games to perform attack-defense analysis and generate defense strategies. 

\subsection{Security Analysis}

\begin{figure}
	\begin{center}
		\includegraphics[scale=0.35]{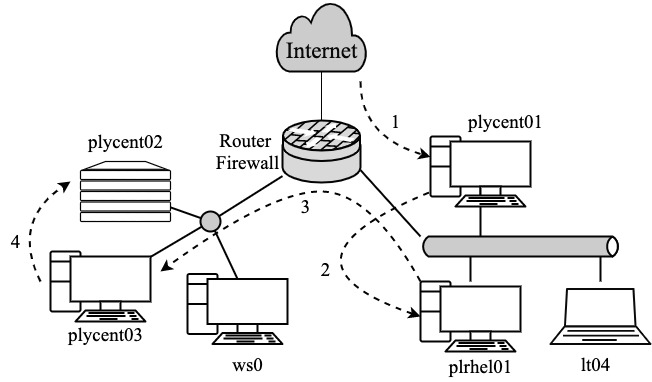}
	\end{center}
	\caption{The network topology and attack steps}
	\label{fig::network.topology}
\end{figure}

\textbf{Case Study Overview.}
\label{subsec:casestudy}
To evaluate our approach, we designed an attack scenario against a network with the topology shown in 
\Figname~\ref{fig::network.topology}. The network consists of three zones: the internet, 
a demilitarized zone (DMZ) that is accessible to the internet and contains the hosts shown to
the right of the router in the figure, and a protected zone that contains the critical services and information in the system. 
This network contains a  router firewall that controls the traffic in the network. 
Each host in the network has a specific purpose: plycent01 is an internet-facing frontend server that acts as an endpoint that exports the services provided by the protected zone to the internet. plyrhel01 acts as a proxy server between the demilitarized and protected zones and is responsible for data and request synchronization and transfer between plycent01 and the application server plycent03. The host plycent02 is the database server that hosts all data required for the system to function.
Table~\ref{tab:servicesandvulns} summarizes
the services and vulnerabilities present of each network host. Table~\ref{tab::vulnerabities} shows the description, and consequences of exploitation for each vulnerability in the system, according to the CVSS database.

The attacker is located in the internet and can access the web pages served by the http server hosted in plycent01. 
S/he first compromises playcent01's web server by exploiting the vulnerability cve-2018-1273, which enables remote code execution on the vulnerable host.
Since remote code execution has been gained on plycent01, the attack continues by gaining root privileges on plycent01 using the vulnerability cve-2017-13215 (Step1).
To gain access to the protected zone, the attacker employs cve-2018-1000120, \ie a buffer overflow vulnerability, in the CURL utility used to coordinate information between plycent01 and plyrhel01, which allows the attacker to obtain the permission of code execution on plyrhel01 (Step2).
plycent3 can then be reached by plyrhel01 via the vulnerable ssh service running on it. cve-2018-7566 is exploited on this vulnerable service to gain root privileges on plycent03 (Step3).
The host plycent02 is required to allow database access to plycen03 in order to be able to store and retrieve information in its database. The PostgresSQL database hosted on it, however, is improperly configured and allows unauthenticated, unregulated access to any user that can connect to it.
Therefore, the attacker can exploit that by performing malicious queries against the database (Step 4).
Malicious database queries can tamper with or delete critical information, or waste time with queries specifically
crafted to do so. A DoS attack can hence be performed in two ways: either by removing the information requested by other parts of the system, or by queuing specifically crafted queries that are irrelevant to the system functional goals.

\begin{table}[ht]
		\caption{Services and Vulnerabilities of the Hosts}
	\centering
	\begin{tabular}{ |c |l|l| }
		\hline
		\textbf{Host} & \textbf{Services} & \textbf{Vulnerabilities } \\ \hline\hline
		\multirow{3}{*}{plyrhel01} & http & cve-2018-1000120  \\
		& http proxy  & cve-2018-1273 \\
		& ssh  & None \\
		
		\hline
		\multirow{2}{*}{plycent01} & http & cve-2018-1273  \\
		& (Linux kernel) &  cve-2017-13215   \\
		\hline
		\multirow{2}{*}{plycent02} & postgress& Improper Authentication  \\
		&ssh & None  \\
		\hline
		\multirow{2}{*}{plycent03} & ssh & cve-2018-7566  \\
		& http & cve-2018-1273  \\
		\hline
	\end{tabular}
	\label{tab:servicesandvulns}
\end{table}
\begin{table*}[ht]
	\centering
	\caption{Description of Vulnerabilities}
	\begin{tabular}{|l|p{170pt}|p{80pt}|p{25pt}|p{25pt}|}
		\hline
		\textbf{Vulnerability} & \textbf{Description} & \textbf{Consequence(s)}& \ \textbf{Exp. Score}& \ \textbf{Imp. Score} \\
		\hline\hline
		cve-2018-1000120 & A buffer overflow exists in CURL in the FTP URL handling. & Remote Code Execution& 3.9 & 5.9\\ 
		\hline
		cve-2017-13215 & Buggy implementation of symmetric key encryption allows control of kernel space code. & Privilege Escalation  & 1.8 & 5.9 \\
		\hline
		cve-2018-1273 & Incorrect sanitization of special characters allows code injection. & Remote Code Execution & 3.9 & 5.9\\
		\hline
		cve-2018-7566 & Buffer overflow in the Linux kernel & Memory Tampering, Authentication Bypass & 1.8& 5.9\\
		\hline
	\end{tabular}
	\label{tab::vulnerabities}
\end{table*}

\textbf{Attacker Behavior.}
In this scenario, the goal is to launch a DoS attack on the web server plycent02.
We set the attack goal to dos(plycent02) and run MulVAL to generate the attack graph showing attack scenarios that can  lead to this target goal. 
The attack graph consists of 50 nodes in total and 18 attack steps.
To obtain the probability of attacker actions, we use the security metric AGP provided by MulVA. This metric expresses the probability that a specific node in the attack graph will be reached
during an attack that is a cumulative score. As an input, we need to provide the base score of each attack step (\ie rule nodes) that states what is the probability of successfully performing an attack step, given that all of its prerequisites are satisfied.

To this end, we use the \emph{exploitability score} provided by CVSS to measure the probability of successful exploitation of a specific vulnerability. This metric considers the four aspects of (a) attack vector (the context by which the exploitation is possible), (b) attack complexity, (c) privileges required to succeed, and (d) whether this attack requires interaction with the user or not.
For the other types of attack steps that don’t exploit a vulnerability and are usually
ordinary activities such as using a service etc, we defer to expert knowledge to set their values properly. In this case study, we set the score for this kind of activities to 0.8, meaning it is more likely the attack step is successful.
Given the base scores, MulVAL generates the AGP metric for each rule node automatically according to the approach introduced in \cite{WangILSJ08}.

We define two reward structures modeling attack/defense costs: (i) \aCosts, that models the costs of an attack step for the attacker, and (ii) \dCosts, that captures the cost of defense in addition to the damages that an attacker can cause to the network due to a successful attack step. 
To estimate the costs in \aCosts, we considered several factors including the time required to perform the attack, ease of exploitation and required tools and requirements  to perform the attack.
To estimate the damage of an attack step (costs imposed to the defender), we use the {impact score} provided by the CVSS database as well as the business security preferences. Impact score quantifies the outcome that an attacker can achieve as a result of exploiting the vulnerability.
Table~\ref{tab::vulnerabities} presents the exploitability score (in the range of [0..4]) and impact scores (in the range of [0..6]) of the vulnerabilities in our case study.



\vskip0.6em

\textbf{Defender behavior.}
The countermeasures that we consider in this attack scenario include stopping a service, limiting access to a service or patching the vulnerabilities.
The costs of such actions depend on many factors, including the criticality of a service, the availability and difficulty of patching the vulnerability, the cost incurred to the organization if the system is not online for a certain period of time, the cost of information leakage or tampering etc.

We considered three cases of systems that can be developed with the topology presented in \Figname~\ref{fig::network.topology}: a tax system, an image hosting system and a hotel booking system.
Each system has its own requirements and preferences in terms of availability of services and the confidentiality and integrity of the information.
For the tax system,  the integrity of information and the availability of the system when in operation is of the utmost importance, and therefore the cost of taking any service down is very high. Patching on the other hand
is of minor importance.
In the image hosting system, availability is important but integrity and confidentiality are not the main concerns.
All three aspects of availability, confidentiality and integrity are associated with some loss of revenue in the hotel booking system, and are then important. 
The cost of a defense varies from 0 to 500.

\textbf{Results.}
In PRISM-games, it is possible to compute the minimum or maximum probability of satisfying a property $\psi$ that can be guaranteed by a coalition $C$  ($\coalition{C} \ms{P_{opt}}~ [\psi]$, $\ms{opt} \in \{min,max\}$).
The maximum or minimum reward can also be computed similarly.
We checked several properties for each game: 

\vskip0.4em
\begin{itemize}
	\item[P1] What is the minimum cost (defense cost and damages caused by the attacker) to defend against the attack and block it: $\coalition{\ms{def}} \ms{R}^{\dCosts}_\ms{min} ~[ \future ~ \attackerBlocked]$.
	
	\item[P2] What is the maximum cost for the attacker to launch a DoS attack against the web server plycent02: $\coalition{\ms{def}} \ms{R}^{\aCosts}_\ms{max} ~[ \future~  \ms{DoS(plycent02)}]$. 
	
	\item[P3] What is the maximum probability that the defender blocks the attack, \ie $$\coalition{\ms{def}} \ms{P_{max}} ~[ \future ~ \attackerBlocked].$$ 
	
	\item[P4] What is the minimum probability that the attacker achieves its final goal, \ie $$\coalition{\ms{att}} \ms{P_{min}}~ [ \future ~ \ms{DoS(plycent02)}].$$ 
\end{itemize}

\vspace{-0.1in}
\vskip0.4em
\begin{table}[ht]
	\centering
	\footnotesize
	\caption{Results}
	\begin{tabular}{|l|l|l|l|l|}
		\hline
		\textbf{Scheduler} &\textbf{Model Size}&\textbf{Property} & \textbf{Case} & \textbf{Result} \\
		\hline\hline
		\multirow{6}{*}{2-actions} &  &
		\multirow{3}{*}{P1} & 
		Tax System & 732 \\\cline{4-5} 
		&&& Image Hosting & 732 \\\cline{4-5} 
		&&& Hotel Reservation & 732 \\\cline{3-5} 
		&17.6K/47k&{P2} & 
		- & 572 \\\cline{3-5} 
		&&	{P3} & 
		- & 0.31 \\\cline{3-5} 
		&&	P4 & - & 0.3 \\\cline{1-5} 
		\multirow{6}{*}{4-actions} & &
		\multirow{3}{*}{P1} & 
		Tax System & 277 \\\cline{4-5} 
		&&& Image Hosting & 266 \\\cline{4-5} 
		&&& Hotel Reservation & 318 \\\cline{3-5} 
		&47K/121k&{P2} & 
		- & 671 \\\cline{3-5} 
		&&	{P3} & 
		- & 0.67 \\\cline{3-5} 
		&&	P4 & - & 0.39 \\\cline{1-5} 
		\multirow{6}{*}{6-actions} & &
		\multirow{3}{*}{P1} & 
		Tax System & 205 \\\cline{4-5} 
		&&& Image Hosting & 185 \\\cline{4-5} 
		&&& Hotel Reservation & 245 \\\cline{3-5} 
		&97K/242K&{P2} & 
		- & 711 \\\cline{3-5} 
		&&	{P3} & 
		- & 0.68 \\\cline{3-5} 
		&&	P4 & - & 0.4 \\\cline{1-5} 
	\end{tabular}
	\label{tab::results}
\end{table}
We carried out experiments with different schedulers.
The attacker performs 14 different types of actions in total and we considered three different schedulers that return the control to the defender after an action of a specific type, e.g., when a suspicious log-in happens or traffic is observed.
Table~\ref{tab::results} shows the analysis results. The first column shows the number of attacker action types after which the control returns back to the defender.
The ``Model Size''' column indicates the number of states and transitions of the security game and ``Case''' shows the corresponding case study.

The analysis results show that, with an increase in the number of actions to which the defender can react, the size of model increases. This is expected, as more transitions from the defender will be included in the model and this will subsequently lead to increasing the state space.
In addition, the minimum cost (attacker damage as well as the defense costs) imposed on the system decreases with an increase in the number of attacker actions to react to, e.g., the minimum cost to defend against the attack in the tax system changes from 732 units to 205 when the number of actions with a defense reaction changes from 2 actions to 6. This is because the defender has a better chance of applying an effective countermeasure if control is returned to it more often.
Furthermore, the minimum cost for the attacker to reach its final goal slightly increases, as with an increase in the defense actions, it might become more challenging for the attacker to reach its goal and need to try different scenarios to succeed.
Similarly, the probability the defender succeeds in blocking the attack increases, e.g.,  from 0.31 with two action types to 0.68 for the case of 6 action types.

When a property is verified by PRISM-games, it can synthesize a strategy that specifies an action to be performed in a state so that the property is satisfied. 
There are 12 types of defense actions in this case study in total. For the property P1 under a 2-action scheduler, 11, 10 and 9 types of actions exist in the strategy for the tax system, the image hosting and the hotel reservation systems, respectively. The strategy synthesized for the tax system includes patching cve-2018-1000120 in plyrhel01, while patching this vulnerability is not recommended for the other two cases.
The vulnerability cve-2018-7566 of the ssh service in plycent03 is not patched in the strategy synthesized for the hotel reservation system.
This could be because blocking ssh traffic from  plyrhel01 to plycent03 is cheaper than patching this vulnerability here. 
We believe that an in-depth interpretation of results requires techniques to synthesize invariants for the states under which a specific action is suggested, and unfortunately this cannot be done manually for games with thousands of states like ours.

\subsection{Performance Analysis}
To evaluate the performance of our approach, we conducted some experiments on a real-life network. This network consists of sixteen hosts offering different services.  We scanned the hosts and identified 155 types of vulnerabilities that are all reported during the period of 2017-2018 to the CVSS database.
All experiments were run on a machine with a 2.9 GHz Intel Core i7 CPU and 16 GB RAM, running MacOS Catalina.
We conducted  experiments with different sets of observation, vulnerabilities, defense strategies and objectives on a real-life case study.  
To test the implementation of our prototype, we ran around 1241 different experiments.

\begin{figure}
	\centering
	\includegraphics[scale=0.6]{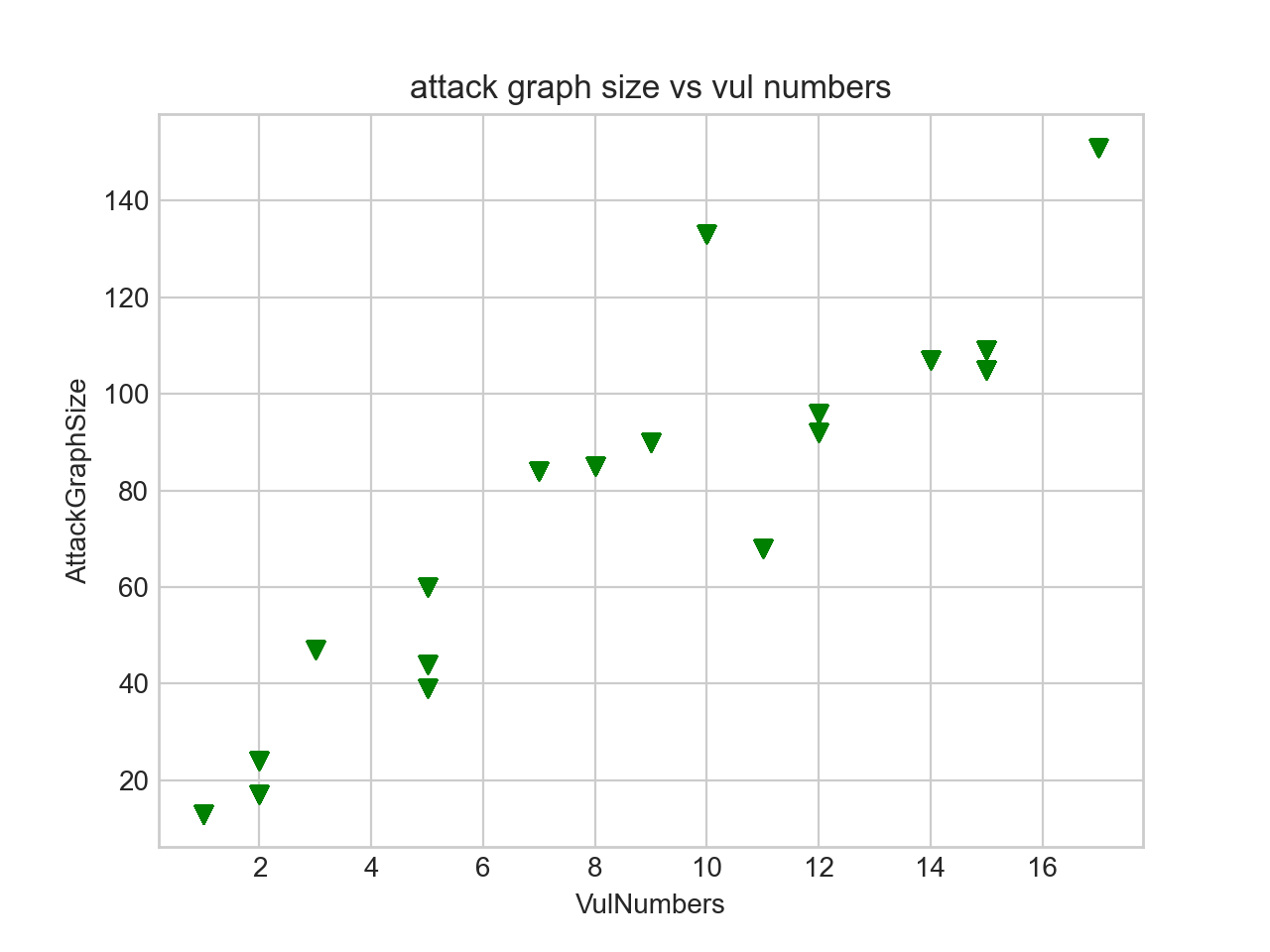}
	\caption{Attack graph size vs number of exploited vulnerabilities.}
	\label{fig:number.of.states.vs.attack.graph.size}
\end{figure}

The size of the game  model depends on the size of the attack graph and the number and complexity of defenses.
The size of attack graph depends on the complexity of attack and the number of potential vulnerabilities exploited to achieve the attack goal, \ie the number of potential attack scenarios and their complexity.
This does not necessarily depend on the size of network, \ie achieving an attack goal in a larger network could be simpler, as it may contain fewer vulnerabilities but which are severe and lead to more privileges being gained in fewer steps. 
As shown in \Figname~\ref{fig:number.of.states.vs.attack.graph.size}, the size of the attack graph  (\ie the number of graph nodes) increases with an increase in the number of vulnerabilities used to achieve a goal.

\begin{figure}
	\centering
	\begin{subfigure}[b]{0.5\textwidth}
		\includegraphics[scale=0.5]{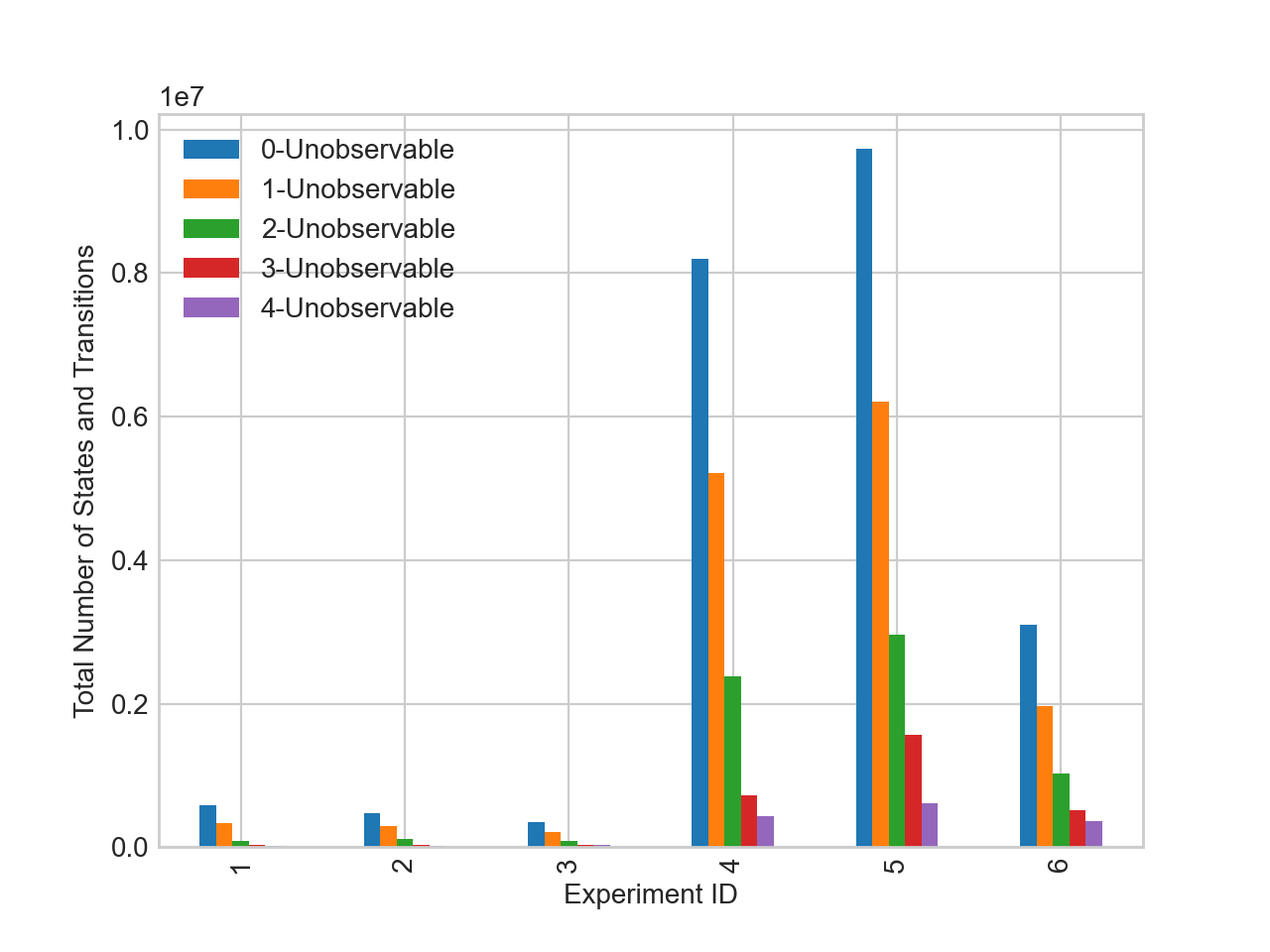}
		\caption{Model Size vs Partial observation}
		\label{fig:partial.observations.results}
	\end{subfigure} 
	\begin{subfigure}[b]{0.45\textwidth}
		\centering
		\includegraphics[scale=0.5]{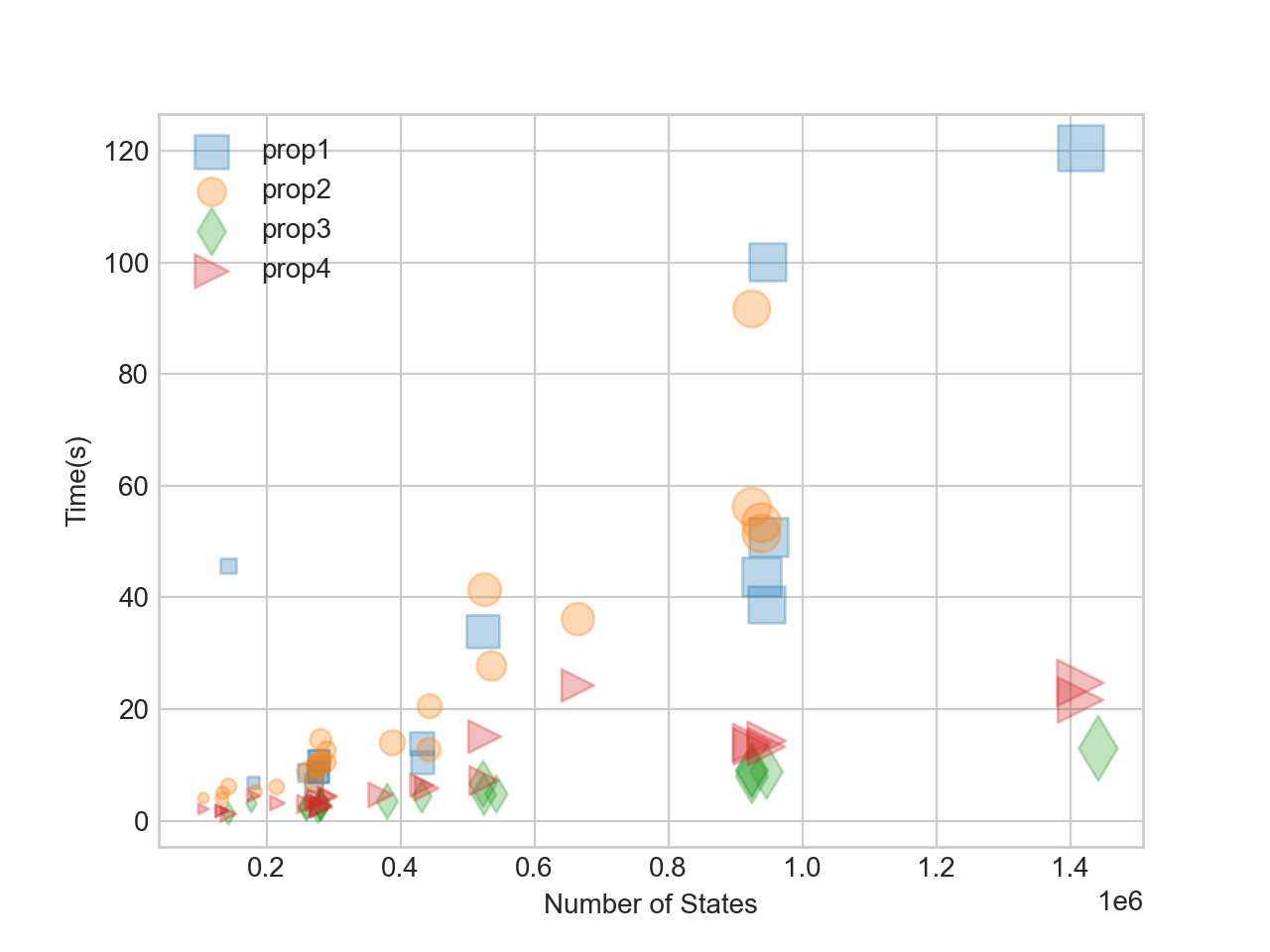}~~~
		\caption{Model Checking Time}
		\label{fig:modeling.checking.time}
	\end{subfigure}	
	\caption{Model size and verification time}
\end{figure}

\begin{figure}
	\centering
	\includegraphics[scale=0.6]{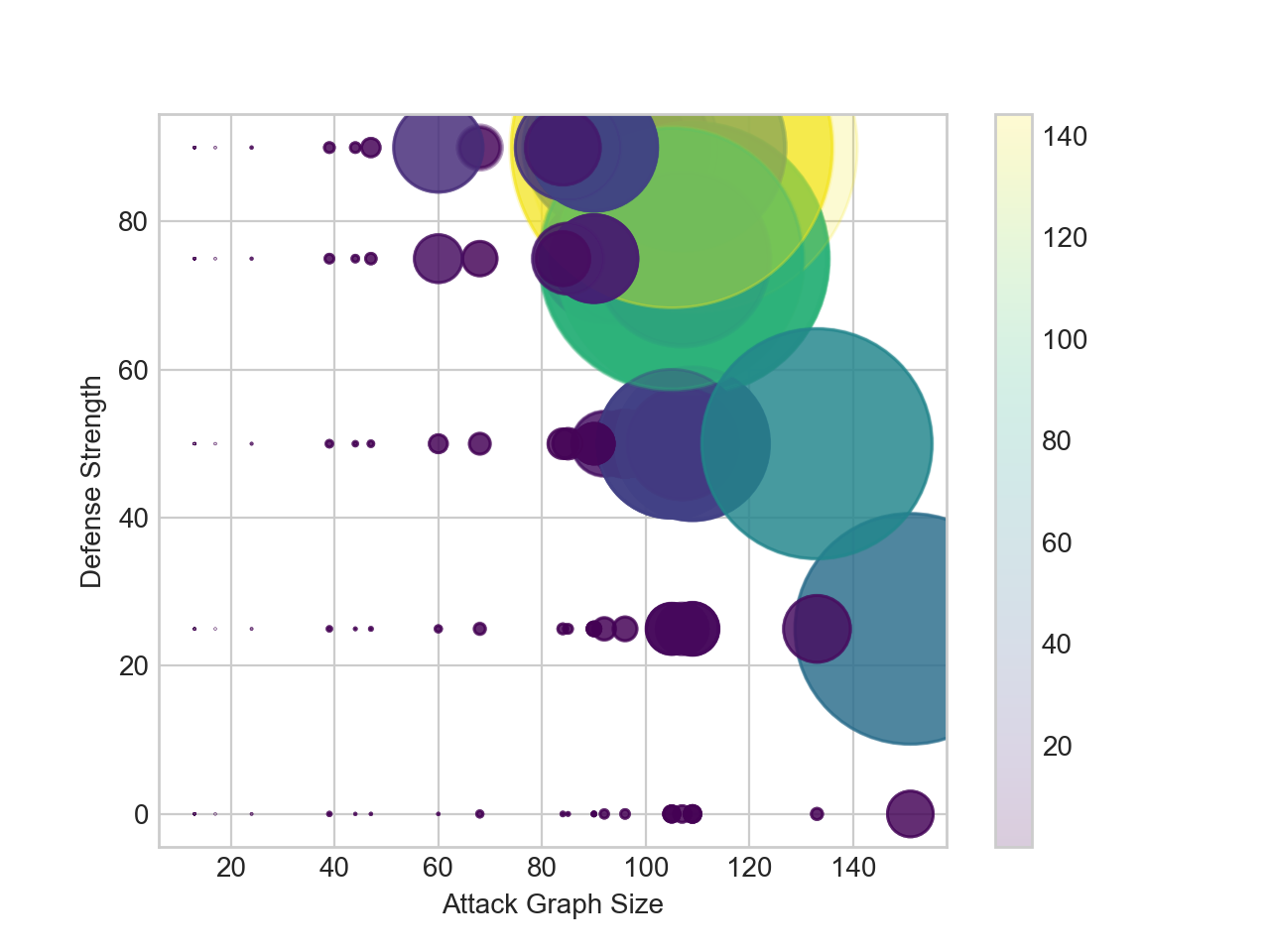}\\
	\scriptsize{
		The color shows the number of states in the interval of $(0,1.4*10^6)$, and the size of point indicates the number of transitions in the interval of $(0,14.5*10^6)$
	}
	\caption{Defense strength, attack graph size, and the game size.}
	\label{fig:number.of.states.vs.defense.game.size}
\end{figure}

Moreover, the size of the game model depends on the size of attack graph as well as the defenses.
\emph{Defense strength} shows the percentage of possible defenses applied in the system. 
With an increase in the size of attack graph and the defense strength, the number of states and transitions increase accordingly, as illustrated in \Figname~\ref{fig:number.of.states.vs.defense.game.size}. 
In this figure, the color is used to show the number of states, and the marker size denotes the number of transitions in the associated point.
For the larger attack graphs with large defense strength, we encountered the state explosion problem.
\Figname~\ref{fig:modeling.checking.time} shows the model checking time of four different properties based on the number of the model states.
The marker size denotes the number of transitions in the corresponding point.
The model checking time was between a few millisecond for the smaller models to  
$\sim$120 seconds for a model with $\sim1.4{\times}10^6$ states,
and often increases with an increase in the size of the model. 

\begin{figure}
	\centering
	\includegraphics[scale=0.5]{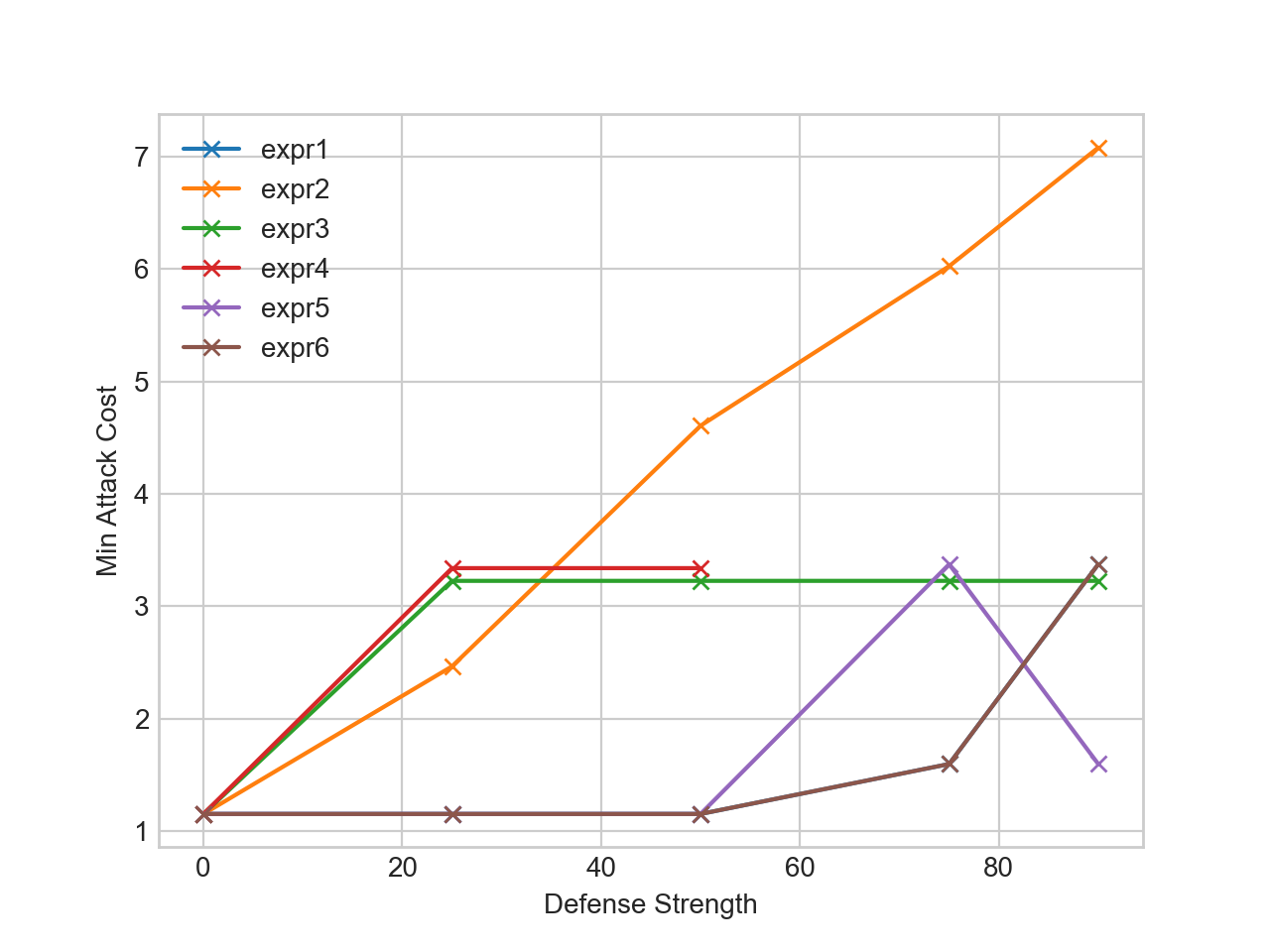}
	\caption{Minimum attack cost vs defense strength.}
	\label{fig:attack.cost.vs.defense.strength}
\end{figure}

In another set of experiments, we studied the effect of the defense strength on the attack cost (\ie the cost for the attacker to successfully achieve its goal).
As shown in \Figname~\ref{fig:attack.cost.vs.defense.strength}, the minimum attack cost often increases  with an increase in the defense strength. Note that the attack cost  depends on other factors such as which defenses are applied, when and where they are applied etc. For instance, the attack cost for the goal expr5 with the defense strength of 90\% is less than that of 75\%.  The reason is that the selected countermeasures in the 90\% case altogether performed weaker compared to the case of 75\%, as they do not block a cheaper attack path prevented by the countermeasures of the case of 75\%. 

The number of transitions of a security game under perfect and partial observation for six different experiments is presented in \Figname~\ref{fig:partial.observations.results}.
The x-axis shows the experiment ID and the y-axis shows the sum of the number of transitions and states for different observations. 
In these experiments, the types of attacker actions was between nine and sixteen types. We run each experiment under five different observations, including full observation (\textsf{Observable}) and $n$ unobservable actions, $1 \leq n \leq 4$.
In this figure, \textsf{$n-$Unobservable} means that $n$ actions of the attacker have become unobservable. 
We considered those actions that are harder to be certainly associated to a malicious activity as unobservable actions.

In some experiments, while the number of state variables remained fixed (because the consequences of unobservable actions were partially observable, \ie neither fully observable nor partially observable), the model size was reduced significantly.  
With an increase in the number of unobservable actions, the number of transitions has been reduced accordingly. For instance, in the experiment 5, the model size for the case of full observation is $\sim$9.7M, while it reduces to $\sim$6.2M, $\sim$2.9M, $\sim$1.5, and 0.6M respectively for 1 to 4 unobservable actions. 
Our experiments show that this transformation can be used as a state reduction technique to tackle the complexity of large-scale attack graphs. We can transform a perfect security game with a large state space to a smaller game, in which some of the sub-scenarios that do not trigger any defense are abstracted away by being considered as a single step. This leads to a more abstract game where the internal details of the attacker's behavior are hidden and only their interactions with the defender are taken into account.

\subsection{Discussion}
Scalability of the approach depends mainly on the number of vulnerabilities that exist in the network and their dependency degree. The more vulnerabilities in the network that can be chained, the larger the game would be. Our results show that it's possible to analyze a real-life network within a reasonable time, however,  the approach may not scale for large networks with dozens of vulnerabilities. A solution to this problem is extending the approach to support modular analysis to handle scalability.

{Furthermore, the verification results are often usable by the user, however, it's not generally always possible for the end user to comprehend synthesis analysis results due to the large state space of games. Our primary goal was to automate defense. Hence, we are integrating our approach in an online monitor with self-protection capabilities to either apply the defenses automatically, or guide the administrator to enforce proper countermeasures at runtime. Accuracy of some parameters such as attack probabilities is an important factor to achieve reasonable results. The online monitor enables us enhance the approach at runtime to improve the accuracy of  parameters based on the attacker's behavior.
}

The assumption that all defense-triggering capabilities of the attacker are observable can be theoretically restrictive, yet a realistic and necessary constraint, as it's impossible to react to an invisible action. If a game ideally needs reacting to an unobservable action, we can modify the game into a similar game with a slightly different defender behavior. One option is to move the defenses of unobservable capabilities to an earlier observable step. This means that the defender should react earlier that can sometimes become more expensive. The other option is to remove such defenses completely, as it's not anyhow feasible to react in such cases. Observe that the results won't be wrong, but they might be less optimal. 

Another limitation of our approach is that it protects the system against known attacks, and zero-day vulnerabilities/attacks cannot be handled by this approach. We plan to enhance it with online attack scenario extraction to be able to handle unknown attacks.
Moreover, the expressiveness of the properties is restricted, as the next and bounded operators are not supported. We believe this subset of rPATL is still reasonable, because we can check important reachability properties, as shown in the experiments.





\section{Related Work}\label{sec:related.work}

\textbf{Games In Security.}
The authors in \cite{DurkotaLBK15} proposed a game-theoretic approach for network hardening problem where the attacker plan is represented as an attack graph and the defender seeks an optimal solution to deploy honeypots into the network to deceive the attacker.
A two-player strategic game based on an extension of attack trees was constructed in~\cite{Bistarelli2007}, where the defender wants to protect the system by buying countermeasures and the attacker tries to exploit the vulnerabilities and gain some profit. The game is solved by finding the Nash equilibria, with the goal of finding the most promising attack/defense's actions. Chowdhary et al. \cite{ChowdharySAHS19} uses attack graphs to build a zero-sum stochastic Markov game between the attacker and the defender to help the admin apply countermeasures in a cloud network.
None of the above approaches supports partial observations.
In contrast to \cite{DurkotaLBK15,ChowdharySAHS19} that are application-specific, our approach can be applied to any type of networks and defenses. Our experiments show that our approach is faster than~\cite{DurkotaLBK15}: we analyze a network with $\sim$53 vulnerabilities in $<1$s while ~\cite{DurkotaLBK15} requires $1s<t<1000s$ for the analysis of a network with  8 types of vulnerabilities.

In \cite{AslanyanNP16}, an extension of attack/defense trees with the temporal ordering of  actions is used to build a two-player stochastic game between the attacker and the defender.  They use probabilistic model checking techniques to formally verify security properties of the attack-defense scenarios, and synthesize strategies for attackers or defenders using PRISM-games.
A similar approach is taken in~\cite{EK19}, but with some relaxed model assumptions 
and some additional optimizations.
Gadyatskay et al. define a stochastic timed semantics for attack-defense trees in terms of a network of timed automata in \cite{GadyatskayaHLLO16}. They use the UPPAAL model checker to perform a quantitative analysis and find attacker parameters that minimize cost.
In contrast to \cite{AslanyanNP16,EK19,GadyatskayaHLLO16}, we use attack graphs that can be generated fully automatically and we support partial observation, while automatic construction of attack trees is still a major challenge despite the recent efforts~\cite{Widel:2019:BFM}. 
Our security game is boolean which makes it more scalable for analysis of realistic network systems, as confirmed by our experiments results.


\vskip0.6em

\textbf{Analysis under Partial Observation.}
Partially observable Markov decision processes (POMDPs) extend MDPs with partial observability and have many applications in fields such as AI, scheduling and planning (\eg \cite{PatilKLSGA14,GradyMK15,JagannathanMMM13,WintererJW0TK017}).  
The formal verification community has studied POMDPs and presented theoretical results, \eg \cite{ChatterjeeCT13,Chatterjee014}, however, progress on developing practical methods has been limited. Norman et al. presented an approach for verification and control of a real-time extension of POMDPs called partially observable probabilistic time automata (POPTAs)~\cite{Norman0Z17}.
Properties are specified using a probabilistic temporal logic and techniques are proposed to either verify the property or to synthesize a controller for the model to enforce it.
Model checking of both POMDPs and POPTAs is implemented in PRISM.

In\cite{WintererJW0TK017}, the problem of motion planning, that is often modeled as a POMDP, is reduced to a two-player stochastic game and solved using PRISM-games.
Also relevant is the recent extension of PRISM-games to concurrent stochastic games~\cite{KNPS18}, which provides an alternative way to model concurrent decision making in a probabilistic setting.
Finally, the problem of controller synthesis for timed games under state-based partial observation has been studied in \cite{CassezDLLR07}
and implemented in UPPAAL-TIGA\cite{behrmann2007uppaal}.
We are not aware of any practical solution for stochastic games under partial observation.


\vskip0.6em

\textbf{Threat Modeling and Analysis.}
Threat modeling and analysis focuses on using models to describe, analyze and discover security problems of a system~\cite{ThreatModelingAG,MulVALPaper,Widel:2019:BFM}. Such techniques are usually provided with quantitative analysis methods to perform security risk assessment, which enables us to understand the attack consequences better and plan more effective countermeasures to defend against attacks. Quantitative analysis in threat models, for the purpose of risk assessment, is often performed based on measuring the security risk of a system using a few security metrics~\cite{8017389}. Attack trees and graphs, including their different variants, are well-established and commonly-used techniques for security modeling. Widel et al.~\cite{Widel:2019:BFM} survey the application of formal methods on interpretation, semi-automatic construction and quantitative analysis of attack trees.
Several variants of attack graphs have been introduced in the literature,  \eg vulnerability cause graphs~\cite{Ardi:2006}, Bayesian attack graph~\cite{Frigault2008} and compromise graphs~\cite{McQueen1579754}. Formal methods have been used to generate attack graphs, \eg Sheyner et al.~\cite{sheyner2002automated,jha2002two} use model checking techniques to generate attack graphs. { Khakpour et al. \cite{KhakpourSNW19} proposed an approach using attack graphs to control the system behavior in a more secure way. } 


%
%

\section{Conclusions}\label{sec:conlusions}
In this paper, we proposed a game-theoretic approach for the analysis of attack graphs under partial observation. We automatically constructed a two-player security game from an attack graph and the defender's behavior  and used PRISM-games to analyze the game and synthesize strategies for the players. To support defense under partial observation, we introduced one-sided partially observable security games and presented an algorithm to transform them to perfect games. We have proven the soundness of our algorithm for ODT security games and a sub-set of objectives. To support the approach, we developed a prototype and conducted some experiments to evaluate it. Our experiments show that the transformation algorithm can be used as a suitable state reduction technique to handle large-scale attack graphs.
 We are currently integrating our approach in a self-protection engine to support online defense. 
We plan to extend our experiments to study the impact of different factors on the effectiveness of a defense.

\emph{\textbf{Acknowledgment}}
{The first author was supported by the Swedish Knowledge Foundation (No. 20160186). }

\bibliographystyle{ACM-Reference-Format}
\bibliography{bibliography}


\vspace{6in}
\clearpage
\pagebreak
\renewcommand\thesection{\arabic{section}}
\renewcommand\thesubsection{\thesection}
\def\thesubsection{\arabic{section}}
\section*{Appendix}\label{sec:soundness-appx}
\newcommand{\oto}[1]{\overset{#1}{\rightarrow}} 
\subsection*{\textsc{A.}~~rPATL Semantics}\label{sec::appendix.rPATL.semantics}
{A path of $\game$ is a possibly infinite sequence $\pi=s_0 \action_0 s_1 \action_1 \ldots$ where ${s_i \in S}$, ${\action_i \in \actions}$, $\gameTrel (s_i,a_i)(s_{i+1}) >0$, for all $0 \leq i (\leq n)$.
	A suffix of $\pi$ starting from the $i$-th state is denoted by $\pi_i= s_ia_i \ldots$.
	A strategy $\strat_i$ for a player ${i \in P}$ (or a coalition of players) is a function ${\strat_i:(S \times \actions)^* S_i \to \Dist_\actions}$ that assigns a probability distribution $\strat_i(\pi.s)$ over actions $\actions(s)$ from a state $s$ with a history of $\pi$~\cite{Sim14}.
	A strategy is memoryless if it does not depend on the history, \ie for all paths $\pi$ and $\pi'$, $\strat_i(\pi.s)=\strat_i(\pi'.s)$.  
	A strategy profile consists of the strategies of all players, \ie $\strat = \strat_1,\strat_2$.
	{By following the strategy $\strat_i$, whenever the history of a game
		played on $\game$ is $\pi.s$ where $s \in S_i$, the player $i$ chooses the next move according to the probability distribution $\strat_i(\pi.s)$.}
	{We denote the set of strategies of the player $i$ by $\strats_i$.}
	The behavior of a game under a given strategy profile is probabilistic,
	and we can define the probability measure $\mathrm{Pr}^{\strat_1,\strat_2}_{\game,s}$
	over its (infinite) paths from a state $s$.
We use $\mathbb{E}^{\strat_1,\strat_2}_{\game,s}(Y)$ to denote the expected value of
a random variable $Y$ with respect to $\mathrm{Pr}^{\strat_1,\strat_2}_{\game,s}$.
}



\begin{definition}[rPATL Semantics \cite{CFK+13b}]\label{def:pATL.semantics}
	The semantics of a state formula $\phi$ is defined inductively for each state $s$ of $\game$ as follows:
	$$
	\begin{array}{rll}
	s \models \top  &   & s \models \alpha  \Leftrightarrow { \alpha \in \slab(s)}\\ 
	s \models \neg \phi &\Leftrightarrow& s \not\models \phi\\ 
	s \models \phi_1 \wedge \phi_2 &\Leftrightarrow& s \models \phi_1 \wedge s \models \phi_2\\ 
	s \models {\coalition{C}} \mathsf{P}_{\bowtie\,p} [\psi] &\Leftrightarrow& \exists \strat_1 \in \strats_{C} \textnormal{ such that } \forall \strat_2 \in \strats_{P \backslash C}\\ &&\mathrm{Pr}^{\strat_1,\strat_2}_{\game,s} (\psi) \bowtie p\\
	s \models {\coalition{C}} \mathsf{R}_{\bowtie\,x} [\future \phi] &\Leftrightarrow& \exists \strat_1 \in \strats_{C} \textnormal{ such that } \forall \strat_2 \in \strats_{P \backslash C}\\ &&\mathbb{E}^{\strat_1,\strat_2}_{\game,s} [rew(r,\future \phi)] \bowtie x\\\\
	\end{array}
	$$
	where $\slab(s)$ returns the set of propositions that hold in the state $s$ and $\mathrm{Pr}^{\strat_1,\strat_2}_{\game,s}(\psi) = 
	\mathrm{Pr}^{\strat_1,\strat_2}_{\game,s}(\{ \pi ~|~ \pi \models \psi  \})$ is a path probability measure  and, informally, gives the probability of satisfying the formula $\psi$ in the state $s$ under the strategy profile $\strat_1,\strat_2$, where, for path $\pi$:
	$$\begin{array}{rll}
	\pi \models \mnext \psi  &\Leftrightarrow& \pi_1 \models \psi\\ 
	\pi \models  \psi_1 \until^{\leq k} \psi_2 &\Leftrightarrow& \pi_i \models \psi_2 \textnormal{ for some } i \leq k \textnormal{ and } \pi_j \models \psi_1 \\
	&&\textnormal{ for } 0 \leq j <i\\ 
	\pi \models  \psi_1 \until \psi_2 &\Leftrightarrow& \psi_1 \until^{\leq k} \psi_2 \textnormal{ for some } k \in \mathbb{N}
	\end{array}
	$$
We use $\mathit{rew}(r,\future \phi)$ to denote the random variable that maps
any infinite path to the sum of rewards $r$ accumulated until a state in $\sat(\phi)$
is reached, or $\infty$ if no such state is reached.
	
\end{definition}

Let ${\opt_s=\max}$ if {$s\in S_C$} (i.e., it's a state of a player from the coalition $C$) and ${\opt_s=\min}$ otherwise.
The probability measure of the next operator is defined as 
$\mathrm{Pr}^{\max,\min}_{\game,s} (\mnext\phi) = \underset{a \in {\actions(s)}}{\opt_s}~ 
\underset{s'\in \sat(\phi)}{\sum} \gameTrel(s,a)s'$.
The probabilities for $\until^{\leq k}$ are defined recursively:

\begin{equation}\label{eqn:pr.of.until.operator}
\mathrm{Pr}^{\max,\min}_{\game,s} (\phi_1 \until^{\leq k} \phi_2) = \begin{cases}
1 & s\in \sat(\phi_2)\\
0 & s\notin (\sat(\phi_1) \cup\sat(\phi_2))\\
0 & k=0,s\in \sat(\phi_1) \backslash \sat(\phi_2)\\
Y & \text{otherwise}
\end{cases}
\end{equation}
where $Y = \underset{a \in{\actions(s)}}{\opt_s}~ 
\underset{s'\in S}{\sum} \gameTrel(s,a)(s') \times \mathrm{Pr}^{\max,\min}_{\game,s'} (\phi_1 \until^{\leq k-1} \phi_2)$.
The unbounded case is computed via value iteration \cite{Condon92}:
\[\mathrm{Pr}^{\max,\min}_{\game,s} (\phi_1 \until \phi_2) = \lim_{k \to \infty} \mathrm{Pr}^{\max,\min}_{\game,s} (\phi_1 \until^{\leq k} \phi_2)\]
Lastly, expected rewards $\mathbb{E}^{\max,\min}_{\game,s}(\future\phi)$
are computed similarly, via value iteration, but using a two-step process
that first computes an upper bound and then converges from above~\cite{CFK+13b}.

The model checking algorithm of rPATL works based on labeling each state $s$ with the formulas that hold in that state, i.e., $\sat(s)$, in a recursive way.
Model checking of properties reduces to computation of optimal
probabilities on the game $\game$. 
For example, if ${\bowtie \in \{ \geq, >\}}$, then
${ s \models {\ll{\{1\}}\gg} \mathsf{P}_{\bowtie~q} [\psi] \Leftrightarrow \mathrm{Pr}^{\max,\min}_{\game,s} (\psi) ~\bowtie~q}$
where $\mathrm{Pr}^{\max,\min}_{\game,s} (\psi) \defop 
\underset{{\strat_1 \in \strats_1}}{\sup}
\underset{{\strat_2 \in \strats_2}}{\inf}\mathrm{Pr}^{\strat_1,\strat_2}_{\game,s} (\psi)$,
i.e., the player $1$ follows the most promising strategies to succeed  while the opponent uses a strategy that is more likely led to failing $\psi$.

\subsection*{\textsc{B.~~}Proof of Theorem~\ref{thm:game.property.verification}}\label{sec::appendix.weak.bisimulation}

Theorem~\ref{thm:game.property.verification} follows from Lemma~\ref{lem:model.check.property.for.any.weak.bisiliar.states} and Theorem~\ref{thm:weak.bisimulation.security.games}.

We use the following notations in this section:
\begin{itemize}
	\item $G$ is an arbitrary attack graph,
	\item $\sattacker= \tuple{\Vars_A, \values_{A,0}, \rulesp, \smdptrans_A}$ defines the semantics of $G$ according to Definition~\ref{def.attack.graph.to.MDP}.
	\item $\actions_{\osattacker} \subseteq \actions_A$ is the set of attacker's observable actions.
	\item $\observable{\sattacker}= \tuple{\Vars_\osattacker, \values_{\osattacker,0}, \actions_{\osattacker}, \smdptrans_\osattacker}$ is a SMDP that specifies the semantics of the {PO}-attacker's behavior under the partial observation $\actions_{\osattacker}$ obtained using Definition~\ref{def:PO.attack.graph.to.MDP}.
	\item $\sdefender = \tuple{\Vars_D, \values_{D,0}, \actions_D, \smdptrans_D}$ is an arbitrary defender.
	\item ${\cattacker}$, $\semantics{\sdefender}$ and $\semantics{\observable{\sattacker}}$ are the MDPs that respectively describe the semantics of $\sattacker$, $\sdefender$ and  $\observable\sattacker$.
	\item $\game = \langle \{A,D\}, V, S_A \cup S_D,\Obs, s_0, \actions,  \Sch, \gameTrel \rangle $ denotes the PO security game $\compGame{{\cattacker}}{\cdefender}{}$ constructed using the functions $\Obs$ and $\sct{up}$ in Equation \ref{eqn:Obs.function} and Equation \ref{eqn:up.function}.
	\item $\observable{\game}= \langle \{A,D\}, \Vars', S'_A \cup S'_D ,  s'_0, \actions', \Sch, \gameTrel' \rangle$ represents the perfect game $\compGame{\observable{\cattacker}}{\cdefender}$.	
	\item $V_{AD} = V_A \cap V_D$ is the set of attacker's defense-triggering capabilities stored in the defender's state, and $V'_{AD}$ is the attacker's copy of the same variables in the PO security game. 
\end{itemize}

We first show that $\game$ and $\observable{\game}$ are weak bisimilar~\cite{Leeuwen90a}.
Let $s =_V s'$ be a notation used to denote that the variables of $V$ have the same values in the states $s$ and $s'$,
i.e., the projections of $s$ and $s'$ on the variables $V$ are identical.
Let $\gameTrel{}$ be a (concrete) transition relation. We define the relation $\multitransrel{}$ as follows:
\begin{itemize}
	\item {$s \multitransrel{\tau} s$,}
	\item $s \multitransrel{\tau} s'$ if $\gameTrel{}(s,a)s' >0$ and $a \notin \actions_{\osattacker}$, and 
	\item $s \multitransrel{\top} s''$ if $s \multitransrel{\tau} s'$, $\gameTrel{}(s',a)s'' >0$ and $a \in \actions_{\osattacker}$.
\end{itemize}

\begin{definition}[Weak Bisimulation]
	Let $\game = \langle \{A,D\}, V,S , s_0, \actions,  \Sch, \gameTrel \rangle$ and $\game' = \langle \{A,D\}, V', S', s'_0, \actions',  \Sch', \gameTrel' \rangle$ be two two-player games.
	A relation $R \subseteq S \times S'$ is a weak bisimulation relation, iff for all $(s,s') \in R$, ${s=_{V \cap V'}s'}$ and  $s \multitransrel{\top} t$ implies that there  exists a state $t' \in S'$ such that $s' \multitransrel{\top} t'$ and 
	$(t,t') \in R$ and vice versa.
	We say $\game$ and $\game'$ are weak bisimilar, denoted by $\game \sim \game'$, if and only if there exists a weak bisimulation relation $R\subseteq S \times S'$ such that $(s_0,s'_0) \in R$.
\end{definition}

\begin{theorem}[Weak Bisimulation of Security Games]\label{thm:weak.bisimulation.security.games}
	The ODT security game ${\game}$ and its transformation to the perfect security game $\observable\game$ are weak bisimilar, i.e., there is a relation $R \subseteq S \times S'$ that is a witnessing weak bisimulation relation where $S = S_A \cup S_D$ and  $S' = S'_A \cup S'_D$..
\end{theorem}
Proof.
To prove this theorem, we should find a relation $R \subseteq S \times S'$ that is a witnessing weak bisimulation relation where $S = S_A \cup S_D$ and  $S' = S'_A \cup S'_D$.
We will show that the following relation is a witnessing weak bisimulation relation and $(s_0,s'_0) \in R$:
\[R= \{ \langle s,s' \rangle~|~ s \downarrow_{(\Vars_{O} \backslash \Vars'_{AD}) \cup \Vars_D \cup \{t\}} = s'   \}. \]
This relation groups the states in which the defender's observations are the same, i.e., the local part of the attacker's state (i.e., $\Vars_{O} \backslash \Vars'_{AD}$), the defender's substate and the scheduler's substates in the ODT security game should be identical to those of the perfect game's state. 

We should show that this candidate relation is a weak bisimulation relation. Consider two states $s$ and $s'$ where $\langle s,s' \rangle \in R$. According to the definition of $R$, the current active player of $s$ and $s'$ should be the same (i.e., $t$ has an identical value in $s$ and $s'$according to $R$). Two cases can happen:

	\textbf{Case i} If $s \in S_D$,  since $\Vars_{AD} \subseteq \Vars_\osattacker$ (from Definition~\ref{def:ODT.security.games}), the same transitions of the defender will be enabled in both games according to Definition~\ref{def:defener.behavior}, Definition~\ref{def:perfect.stochastic.game}, Definition~\ref{def:PO.Security.Games} and Equation~\ref{eqn:Obs.function}. Since the defender's state in $s$ and $s'$ are identical according to $R$, the updates by the defender (i.e., to the variables $V_D$ and $V_\osattacker$) will be the same according to Equation~\ref{eqn:up.function}, Definition~\ref{def:perfect.stochastic.game} and Definition~\ref{def:PO.Security.Games}, i.e., the target states $t$ and $t'$ will be in the relation $R$ too.

		\textbf{Case ii} If $s \in S_A$,  let $s \multitransrel{\top} \gamma$, i.e., $s \oto{\action_0} s_0 \ldots s_{n}\oto{\action_n} \gamma $ where $\sigma_n \in \actions_\osattacker$ and $\sigma_i \notin \actions_\osattacker$, $0 \leq i <n$. 
	According to Definition~\ref{def.attack.graph.to.MDP}, an attacker tries an action, if it leads to gaining a new capability. This means that this sequence leads to updating the consequences of each one of the actions in the sequence, i.e., $e_i \mapsto \top, 0 \leq i \leq n$ where $e_i$ is the consequence capability of $\sigma_i$. According to Equation~\ref{eqn:up.function}, $e_i$'s will be updated in the attacker's substate of $\gamma$ where $0 \leq i \leq n$, while only $e_n$ is updated in the $\gamma$'s defender's substate. 

	It is trivial to show that this sequence corresponds to a transition with a
	 
	 $\sct{dop}= \underset{c_i \in \pred(\sigma_i), 0 \leq i \leq n, c_i \notin \{e_0, \ldots, e_n\} }{\bigcup}c_i$ 
	 in the perfect game $\observable{\game}$ according to Definition~\ref{def:PO.attack.graph.to.MDP} and Definition~\ref{def:perfect.stochastic.game}. This transition will lead to a state $\gamma'$ in which $e_n$ will be set, i.e., $e_n \mapsto \top$. Hence, the projection of $\gamma$ on the union of the attacker's local substate, the defender's state and the scheduler (according to Equation~\ref{eqn:PO.scheduler.function}) will be identical to $\gamma'$, i.e., $\gamma \downarrow_{(\Vars_{O} \backslash \Vars_{AD}) \cup \Vars_D \cup \{t\}} = \gamma'$ and $\langle \gamma, \gamma' \rangle \in R$. The other direction is proven similarly.
	From Definition~\ref{def.attack.graph.to.MDP}, Definition~\ref{def:PO.attack.graph.to.MDP}, Definition~\ref{def:perfect.stochastic.game}, Definition~\ref{def:PO.Security.Games} and Equation~\ref{eqn:Obs.function}, it is concluded that the defender's observations are identical in the initial states of both games, i.e., $\langle s_0,s'_0 \rangle \in R$.
%

%

Since a two-player security game is a zero-sum game, i.e., the players have opposite objectives, an objective can be specified in terms of a single-player strategies and ${\bowtie \in \{ \geq, >\}}$ using the equivalences defined on rPATL formulas \cite{Sim14}.
Hence, we focus on the maximization strategies in the proof, i.e., $\bowtie \in \{ >, \geq\}$.
\begin{lemma}\label{lemm:formula.preservance.in.hidden.path}
	Let $\pi = s_1 a_1 s_2 \ldots a_{n-1} s_n$ be an attacker's \emph{maximal unobservable path} in an ODT security game, i.e., $s_1 \multitransrel{\top} s_n$.
	Let $\phi$ be an observable step-unbounded defense objective. 
	The formula $s_j \in \sat(\phi)$ implies $s_{j+1} \in \sat(\phi)$, for all $1 \leq j < n-1$.
\end{lemma}
Proof.	The proof is done by structural induction on $\phi$.
	\\\textbf{Base Case}
	According to Definition~\ref{def.attack.graph.to.MDP}, each transition by the attacker leads to setting a derived capability,
	i.e., it only gains a new unobservable capability $e_j$ in each transition. Hence,  
	\begin{equation}\label{eqn:lem:Pr.formula.of.hidden.path.state.labels}
	\slab(s_{j+1}) = \slab(s_j) \cup \{e_j\}, 1 \leq j <n-1.
	\end{equation}
	where $e_j \in \Vars_{A} \backslash \Vars_{O}$.
	Since $\phi$ is a proposition defined over $\Vars_{O} \cup \Vars_{D}$ according to Definition~\ref{def:assumptions}, it follows that $s_j \in \sat(\phi)$ implies $s_{j+1} \in \sat(\phi)$, $1 \leq j \leq n$ based on Equation~\ref{eqn:lem:Pr.formula.of.hidden.path.state.labels}.
	\\\textbf{Inductive Step}. If the formula is of the form $\neg \phi$ or $\phi_1 \wedge \phi_2$, the conclusion is trivially followed from the inductive step hypothesis {and the fact that if two formulas hold in a state, their conjunction and negations will hold too.}
	If it's of the form ${\ll C\gg \mathsf{P}_{\leq~q} [\phi_1 \until \phi_2]}$, from the monotonocity property of the value iteration~\cite{Condon92}, the formula
		$ \mathrm{Pr}^{\max,\min}_{\game,s_j}(\phi_1 \until \phi_2) \leq \mathrm{Pr}^{\max,\min}_{\game,s_{j+1}}(\phi_1 \until \phi_2)$ holds.
		Hence,  $ q \leq \mathrm{Pr}^{\max,\min}_{\game,s_j}(\phi_1 \until \phi_2)$ implies $ q \leq \mathrm{Pr}^{\max,\min}_{\game,s_{j+1}}(\phi_1 \until \phi_2)$, and subsequently according to the semantics of rPATL, $s_j \in \sat({\ll C\gg \mathsf{P}_{\leq~q} [\phi_1 \until \phi_2]})$ implies $s_{j+1} \in \sat{{(\ll C\gg \mathsf{P}_{\leq~q} [\phi_1 \until \phi_2]})}$.

\begin{lemma}\label{lemm:Pr.formula.of.hidden.path}
	Let $\Pi$ show the set of all {attacker's maximal unobservable paths} from ${s_1 \in S_A}$, $\phi_1$ and $\phi_2$ be two arbitrary observable step-unbounded defense formulas and $s_1 \in \sat(\phi_1)$ and $s_1 \notin \sat(\phi_2)$.
	It holds ${ \mathrm{Pr}^{\max,\min}_{\game,s_1}(\phi_1 \until \phi_2)=
		\underset{\pi \in \Pi, \pi=s_1 a_1 s_2 \ldots a_{n-1}s_n}{\min} ~\overset{\Rightarrow}{\gameTrel}(\pi)}$ where  
	$$ \begin{array}{ll}
	\overset{\Rightarrow}{\gameTrel}(\pi) =& \gameTrel(s_1,a_1)(s_2) \times \ldots \times \gameTrel(s_{n-1},a_{n-1})s_n \times \\ &\mathrm{Pr}^{\max,\min}_{\game,s_n}(\phi_1 \until \phi_2).
	\end{array}$$
\end{lemma}

Proof.
	From Lemma~\ref{lemm:formula.preservance.in.hidden.path}, it follows that $s_1 \in \sat(\phi_1)$ implies $s_j \in \sat(\phi_1)$, $1<j \leq n$ for all the traces $\pi \in \Pi$.
	Hence, since $\phi_1 \until \phi_2$ is an observable step-unbounded defense objective formula (i.e., it does not change defender's variables), it is trivial to show that the fourth case of Equation~\ref{eqn:pr.of.until.operator} will hold for all the states $s_j$ (excluding their last states) of the $\Pi$'s traces (with $k = \infty$), $1<j \leq n$ and for all traces $\pi \in \Pi$.
	{From the hypothesis $\Vars_{AD} \subseteq \Vars_{O}$ and Definition~\ref{def.attack.graph.to.MDP} (the attacker gains a new capability in each attempt), we can conclude that there will be no loop in the unobservable states, i.e., all traces in $\Pi$ will be finite. 
		Therefore, the conclusion is obviously followed from the definition of Equation~\ref{eqn:pr.of.until.operator} where $opt=\min$. 
	} 

%
%

\begin{lemma}\label{lem:model.check.property.for.any.weak.bisiliar.states}
	Let $\phi$ be an observable step-unbounded defense objective.
	For any arbitrary states $s \in S_A \cup S_D$ and $s' \in S'_A \cup S'_D$ that are weak bisimilar, i.e., $(s,s') \in R$ where $R$ is a witnessing bisimulation relation between $\game$ and $[\game]$. The formula $\game, s \models \phi$ holds if and only if $[\game],s' \models \phi$ holds.
\end{lemma}

Proof.
	We prove this lemma by induction on the number of path operators in $\phi$.
	\\\textbf{Base Case}  The formula $\phi$ is a propositional formula on $\Vars_{O} \cup \Vars_{D}$. From the definition of the witnessing relation (See the proof of Theorem~\ref{thm:weak.bisimulation.security.games}), we can conclude that $\phi \in \slab(s)$ if and only if $\phi \in \slab(s')$, i.e., $\game, s \models \phi$ holds if and only if $[\game],s' \models \phi$ holds.
	\\\textbf{Inductive Step} Assume that the conclusion holds for all the formulas $\phi_i$ with $i$ operators of $\mathsf{P}$, $0 \leq i \leq n$.
	We should prove it for $n+1$, i.e., it should hold for all formulas $\phi_{m}$ with $m$ operators of $\mathsf{P}$, $0 \leq m \leq n+1$. 
	If $0 \leq m \leq n$, the conclusion follows from the inductive step hypothesis. 
	For the case of $m=n+1$, we perform a case analysis on $\phi_{m}$ (i.e., $\phi_{n+1}$). (\textbf{Case a}) If $\phi_m=\neg \phi$, the proof is trivial. 
	(\textbf{Case b}) If $\phi_m=\phi_{i} \wedge \phi_j$ where $i+j=n+1$, (i) if $i\neq 0$ and $j \neq 0$, it trivially follows from the inductive step hypothesis and the fact that if two formulas hold in a state, their conjunction holds as well, and (i) if $i=0$ or $j=0$, then the proof is reduced to proving ${\ll{C}\gg} \mathsf{P}_{\bowtie~q} [\psi]$ or $ {\coalition{C}} \mathsf{R}_{\bowtie\,x} [\future \phi]$ similar to Case c and d.
		
	(\textbf{Case c}) If ${\ll{C}\gg} \mathsf{P}_{\bowtie~q} [\psi]$, since this formula has $n+1$ number of path operators, it follows that the formula $\psi$ should have $n$ path operators. As $\psi$ is an observable step-unbounded defense objective, $ \psi = \phi_1 \until \phi_2$. 
	We first prove the left-to-right direction of this equivalence, i.e., ${\game, s \models \phi \implies [\game],s' \models \phi}$.
	Since $\game,s \models {\ll{C}\gg} \mathsf{P}_{\bowtie~q} [\psi]$, this means that there is at least a path $\pi=sa_0 s_0a_1 \ldots s_n$ in $\game$ that { follows a winning strategy and leads in satisfying this property, i.e., $ \mathrm{Pr}^{\max,\min}_{\game,s} (\psi) = \mathrm{Pr}^{\max,\min}_{\game,s} (\{\pi\})$, $q \bowtie \mathrm{Pr}^{\max,\min}_{\game,s} (\{\pi\})$ and $ \mathrm{Pr}^{\max,\min}_{\game,s} (\psi) = \mathrm{Pr}^{\max,\min}_{[\game],s'} (\psi)$.} We should show that there is a path $\pi'$ in $[\game]$ such that  $ \mathrm{Pr}^{\max,\min}_{[\game],s'} (\psi) = \mathrm{Pr}^{\max,\min}_{[\game],s'} (\{\pi'\})$, ${(s,s') \in R}$ and $q \bowtie \mathrm{Pr}^{\max,\min}_{[\game],s'} (\psi)$.
		We prove this by induction on the number of $a_i$'s in $\pi$.
		\begin{itemize}
			\item \textbf{Base Case} If $\pi = s$, we show that $\pi'=s'$. {According to the first inductive step hypothesis,  $s =_{\Vars_{O} \cup \Vars_{D}} s'$ (witnessing bisimulation relation) and Lemma~\ref{lemm:formula.preservance.in.hidden.path}}, one of the first three cases in Equation~\ref{eqn:pr.of.until.operator} should hold in both $s$ and $s'$. Hence, $\mathrm{Pr}^{\max,\min}_{\game,s} (\psi) = \mathrm{Pr}^{\max,\min}_{[\game],s'} (\psi)$ and subsequently $ [\game],s' \models \phi$ holds.
			
			\item \textbf{Inductive Step} Let for all paths $\pi_k$ with the $k$ number of actions where $\mathrm{Pr}^{\max,\min}_{\game,s} (\psi) = \mathrm{Pr}^{\max,\min}_{\game,s} (\{\pi\})$, there is a path $\pi'$ such that $\mathrm{Pr}^{\max,\min}_{[\game],s'} (\psi)= \mathrm{Pr}^{\max,\min}_{[\game],s'} (\{\pi'\})$, $q \bowtie \mathrm{Pr}^{\max,\min}_{[\game],s'} (\psi)$ and $ [\game],s' \models \phi$ holds. 
			We should prove this for $\pi_{k+1}=s a_0 s_0 a_1 \ldots s_{k+1}$.
			Two cases can happen. 
			
			 \textbf{Case i} If $s \in S_D$: Let $\pi_0= s_0 a_1 \ldots s_{k+1}$. Based on Equation~\ref{eqn:pr.of.until.operator} and ${0<k}$, it is clear that $\mathrm{Pr}^{\max,\min}_{\game,s_0} (\psi)= \mathrm{Pr}^{\max,\min}_{\game,s_0} (\{\pi_0\})$, $q \bowtie \mathrm{Pr}^{\max,\min}_{\game,s_0} (\psi)$ and $ \game,s_0 \models \phi$.
				From the inductive step hypothesis, there is a path $\pi_0'=s'_0 a'_1 \ldots s'_{j}$ such that $(s_0,s'_0) \in R$, ${q \bowtie \mathrm{Pr}^{\max,\min}_{[\game],s'_0} (\psi)}$ and $[\game],s'_0 \models \phi$.				
				From the definition of witnessing bisimulation relation and the disjointness of the attacker and the defender's actions, it follows $s \in S_D \Leftrightarrow s' \in S'_D$ where $(s,s') \in R$. According to the first inductive step  hypothesis, $s \in \sat(\phi_1)$ if and only if $s' \in \sat(\phi_1)$. Similarly for $\phi_2$. 
				Hence, the same case of Equation~\ref{eqn:pr.of.until.operator} holds for the states of $s$ and $s'$, i.e.,  the fourth case. 
				The outgoing transitions from $s$ and $s'$ are the same in the both games, as the defender is same in both games. Hence, we can conclude from the inductive step hypothesis ($\mathrm{Pr}^{\max,\min}_{\game,s_0} (\psi) = \mathrm{Pr}^{\max,\min}_{[\game],s'_0} (\psi)$), the definition of weak bisimulation relation and Equation~\ref{eqn:pr.of.until.operator} that $\mathrm{Pr}^{\max,\min}_{\game,s} (\psi) = {\mathrm{Pr}^{\max,\min}_{[\game],s'} (\psi) }$, i.e., $ [\game],s' \models \phi$
				
			 \textbf{Case ii}	If ${s \in S_A}$: 
				According to Lemma~\ref{lemm:Pr.formula.of.hidden.path}, there is a path $\omega_0=s_0 a_0 s_1 \ldots a_{j-1}s_j$ such that   $\mathrm{Pr}^{\max,\min}_{\game,s} (\psi) = \underset{0\leq i < j}{\Pi} \gameTrel(s_i,a_i)s_{i+1} \times \mathrm{Pr}^{\max,\min}_{\game,s_j} (\psi)$, 
				i.e., $\pi_{k+1}=\omega_0 \omega_1$ and $\mathrm{Pr}^{\max,\min}_{\game,s_j} (\psi) = \mathrm{Pr}^{\max,\min}_{\game,s_j} (\{\omega_1\})$.
				From the inductive step hypothesis, it follows that there is a path $\omega'_1=s'_j a_j \ldots$ such that $\mathrm{Pr}^{\max,\min}_{[\game],s'_j} (\psi) = \mathrm{Pr}^{\max,\min}_{\game,s_j} (\{\omega'\})$.
				Since ${(s_0,s'_0) \in R}$,  $\pi' = s'_0a \omega'_1$ is a path of $[\game]$.
				According to Algorithm~\ref{alg:attack.graph.optimization}, it follows that 
				$\gameTrel(s'_0,a)s'_j = \underset{0\leq i < j}{\Pi} \gameTrel(s_i,a_i)s_{i+1}$, i.e., $\mathrm{Pr}^{\max,\min}_{\game,s_0} (\{\pi\})  = \mathrm{Pr}^{\max,\min}_{[\game],s'_0} (\{\pi'\})$, and subsequently 
				$\mathrm{Pr}^{\max,\min}_{\game,s_0} (\psi) = \mathrm{Pr}^{\max,\min}_{[\game],s'_0} (\psi)$ and 
				$ [\game],s_0' \models \phi$. 
			\end{itemize}
		The proof from right-to-left is done similarly.			
		
		(\textbf{Case d}) If $ {\coalition{C}} \mathsf{R}_{\bowtie\,x} [\future \phi]$, the proof is similar to that of  $ {\coalition{C}} \mathsf{P}_{\bowtie\,x} [\psi]$. Note that the semantics of the $\mathsf{R}$ operator is defined based on $\mathrm{Pr}^{\strat_1,\strat_2}_{\game,s}$ and the accumulated reward of a path. 
		
		

\subsection*{\textsc{C.~~}Proof of Theorem~\ref{thm:game.property.synthesis}}\label{sec::appendix.synthesis}

Proof.	A wining strategy is a function $\strat: S_D \to \dist_S$ that determines the next move in each defender's state $s$ based on its observation.
	Let $T=\{t \in \gameTrel(s)~|~ \mathrm{Pr}^{\max,\min}_{\game,s} (\psi) \leq \mathrm{Pr}^{\max,\min}_{\game,t} (\psi)\}$.
	Then, $\strat(s)=[t_1 \mapsto 1/n, t_2 \mapsto 1/n, \ldots, t_n \mapsto 1/n]$ 
	if $T\neq \emptyset$,
	otherwise, 
	$\strat(s)=[s_1 \mapsto 1/m, s_2 \mapsto 1/m, \ldots, s_m \mapsto 1/m]$ where $\gameTrel(s) = \{ s_1, \ldots, s_m\}$.
	Informally, as long as the property holds, the defender will take the most promising states that more likely lead to holding the property $\psi$. As soon as the property is violated or satisfied, then it can randomly move.
	As shown in Theorem~\ref{thm:game.property.verification}, for any state $s\in S_D$, $\mathrm{Pr}^{\max,\min}_{\game,s} (\psi) = \mathrm{Pr}^{\max,\min}_{[\game],s'} (\psi)$ where $(s,s') \in R$.
	Hence, we can conclude that the synthesized strategy will be identical in both cases, if the property is of the form ${\ll{C}\gg} \mathsf{P}_{\bowtie~q} [\psi]$.
	The proof for the case of ${\ll{C}\gg} \mathsf{R}_{\bowtie~q} [\future \phi]$ is similar.

\end{document}